\documentclass{ws-ijmpe}

\def\lsim{\raise0.3ex\hbox{$<$\kern-0.75em\raise-1.1ex\hbox{$\sim$}}}
\def\gsim{\raise0.3ex\hbox{$>$\kern-0.75em\raise-1.1ex\hbox{$\sim$}}}

\def\mean#1{\left<#1\right>}
\def\Journal#1#2#3#4{{#1}{\bf#2} (#4) #3}

\def\IJMPE{{\it Int. J. Mod. Phys.}~{\bf E}}
\def\EPJC{{\it Eur. Phys. J.}~{\bf C}}

\def\JHEP{{\it JHEP\ }}
\def\JPG{{\it J. Phys.}~{\bf G}}

\def\NIMA{{\it Nucl. Instrum. Methods}~{\bf A}}
\def\NPA{{\it Nucl. Phys.}~{\bf A}}
\def\NPB{{\it Nucl. Phys.}~{\bf B}}
\def\PLB{{\it Phys. Lett.}~{\bf B}}
\def\PLC{\it Phys. Repts.\ }
\def\PRL{\it Phys. Rev. Lett.\ }
\def\PRD{{\it Phys. Rev.}~{\bf D}}
\def\PRC{{\it Phys. Rev.}~{\bf C}}

\def\PPNP{{\it Prog. Part. Nucl. Phys.}}
\def\ZPC{{\it Z. Phys.}~{\bf C}}
\def\ARNS{{\it Ann. Rev. Nucl. Part. Sci.\ }} 

\def\RPP{\it Rep. Prog. Phys.\ }

\begin{document}

\markboth{M. J. TANNENBAUM}{HEAVY ION PHYSICS AT RHIC}

\catchline{}{}{}{}{}

\title{HEAVY ION PHYSICS AT RHIC\footnote{Research supported by U. S. Department of Energy, DE-AC02-98CH10886.}
}

\author{\footnotesize M. J. TANNENBAUM}

\address{Physics Department, 510c,\\
Brookhaven National Laboratory,\\
Upton, NY 11973-5000, USA\\
mjt@bnl.gov}\maketitle

\begin{history}
\received{(received date)}
\revised{(revised date)}
\end{history}

\begin{abstract}
The status of the physics of heavy ion collisions is reviewed based on measurements over the past 6 years from the Relativistic Heavy Ion Collider (RHIC) at Brookhaven National Laboratory. The dense nuclear matter produced in Au+Au collisions with nucleon-nucleon c.m. energy $\sqrt{s_{NN}}=200$ GeV at RHIC corresponds roughly to the density and temperature of the universe a few microseconds after the `big-bang' and has been described as ``a perfect liquid" of quarks and gluons, rather than the gas of free quarks and gluons, ``the quark-gluon plasma" as originally envisaged. The measurements and arguments leading to this description will be presented.   
\end{abstract}

\section{Introduction}\label{sec:introduction}
High energy nucleus-nucleus collisions provide the means of creating nuclear matter in conditions of extreme temperature and density~\cite{BearMountain,seeMJTROP,MJTROP}.  
 The kinetic energy of the incident projectiles would be dissipated in the large 
volume of nuclear matter involved in the reaction.  The system is expected 
to come to equilibrium, thus heating and compressing the nuclear matter so that it undergoes a phase transition from a state of nucleons containing bound quarks and gluons to a state of deconfined quarks and gluons, in chemical and thermal equilibrium, covering the entire
volume of the colliding nuclei or a volume that corresponds to many units of the characteristic length scale. This state of nuclear matter was originally given the name Quark Gluon Plasma(QGP)~\cite{Shuryak80}, a plasma being an ionized gas. However the results at RHIC to be presented here indicated that instead of behaving like a gas of free quarks and gluons, the matter created in heavy ion collisions at nucleon-nucleon c.m. energy $\sqrt{s_{NN}}=200$ GeV appears to be more like a {\em liquid}~\cite{seeMJTROP}. This matter interacts much more strongly than originally expected, as elaborated in recent peer reviewed articles by the 4 RHIC experiments~\cite{BRWP,PHWP,STWP,PXWP}, which inspired the theorists~\cite{THWPs} to give it the new name ``sQGP" (strongly interacting QGP).  

	Two energy regimes are discussed for the QGP~\cite{Anishetty80}. 
At lower energies, $\sqrt{s_{NN}}\simeq 5-10$ GeV, typical of the AGS and CERN  fixed target programs~\cite{seeMJTROP}, the colliding nuclei are 
expected to stop each other, leading to a baryon-rich system. This 
will be the region of maximum baryon density. At very high energy,
$\sqrt{s_{NN}}\ \gsim 100 - 200$ GeV, nuclei become transparent and the nuclear
fragments will be well separated from a central region of particle
production at mid-rapidity. This is the region of the baryon-free or gluon plasma, while in the nuclear fragmentation regions a baryon-rich plasma may also be formed~\cite{Anishetty80}. 

	In the terminology of high energy physics, 
the QGP or sQGP is called a ``soft'' process, related to the QCD confinement scale 
\begin{equation}
\Lambda^{-1}_{\rm QCD} \simeq {\rm (0.2\ GeV)}^{-1} \simeq 1 \, 
\mbox{fm}\qquad .
\label{eq:LambdaQCD}
\end{equation}
   With increasing temperature, $T$, in analogy to increasing $Q^2$, the strong coupling constant $\alpha_{s}(T)$ becomes smaller, reducing the binding energy,  and the string tension, $\sigma(T)$, becomes smaller, increasing the confining radius, effectively screening the potential\cite{SatzRPP63}: 
  \begin{equation}
  V(r)=-{4\over 3}{\alpha_{s}\over r}+\sigma\,r \rightarrow 
-{4\over 3}{\alpha_{s}\over r} e^{-\mu\,r}+\sigma\,{{(1-e^{-\mu\,r})}\over \mu}
\label{eq:VrT}
\end{equation} 
where $\mu=\mu(T)=1/r_D$ is the Debye screening mass~\cite{SatzRPP63}. For $r< 1/\mu$ a quark feels the full color charge, but for $r>1/\mu$, the quark is free of the potential, effectively deconfined. 

\begin{figure}[!thb]
\begin{center}
\begin{tabular}{cc}
\psfig{file=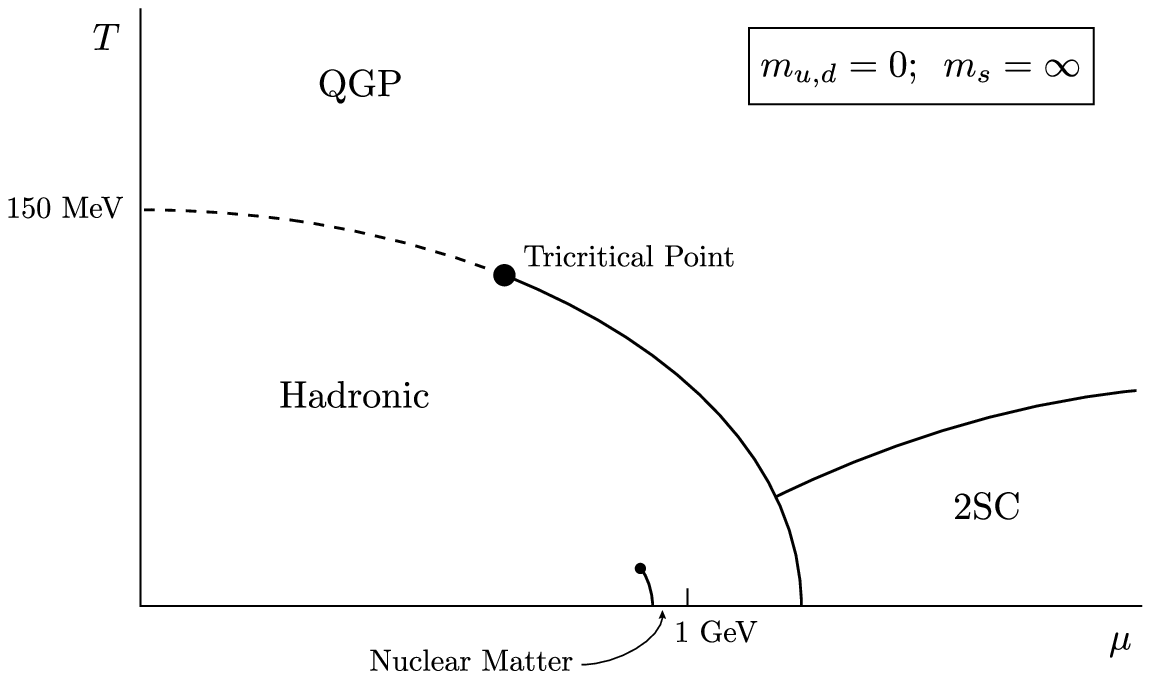,width=0.49\linewidth} &\hspace*{-0.03\linewidth}
\psfig{file=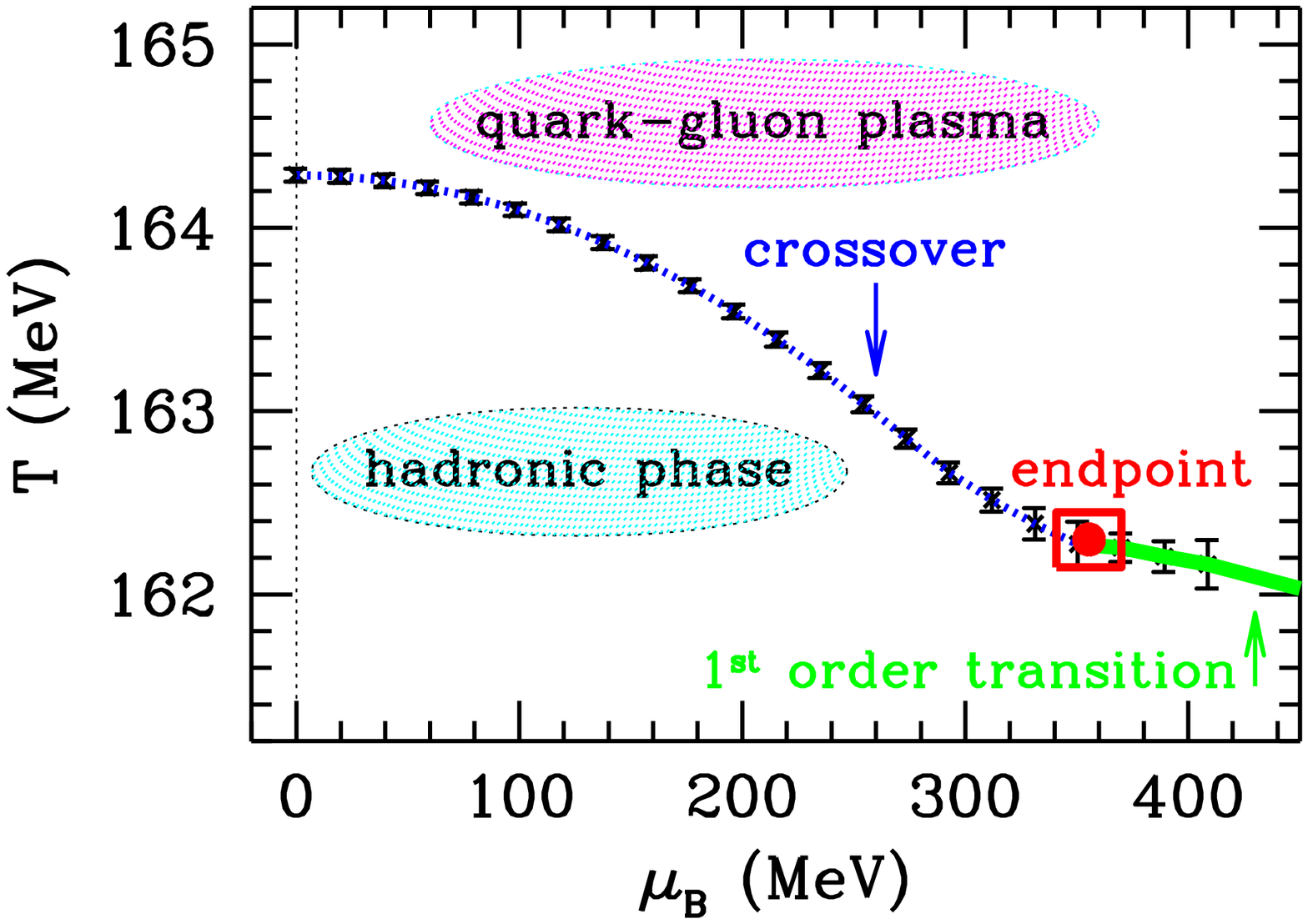,width=0.49\linewidth} 

\end{tabular}
\end{center}
\caption[]{a)(left) Proposed phase diagram for nuclear matter~\cite{Krishna99}: Temperature,  $T$,  vs Baryon Chemical Potential, $\mu$. b) (right) Lattice calculation with ``2+1 staggered quarks of physical masses''~\cite{FodorKatz}. \label{fig:phaselat}}

\end{figure}

   There has been considerable work over the past 
three decades in making quantitative predictions for the QGP~\cite{seeMJTROP}. The predicted transition temperature from a state of hadrons to the QGP varies, from $T_c\sim 150$ MeV at zero baryon density, to zero temperature at a critical baryon density roughly 1 GeV/fm$^3$, 
$\sim$ 6.5 times the normal density of cold nuclear matter,  
$\rho_0 = 0.14\,  {\rm nucleons}/ {\rm fm}^3$, $\mu_B\simeq 930$ MeV, 
where $\mu_B$ is the Baryon chemical potential. A typical expected phase diagram of nuclear matter~\cite{Krishna99} is shown in Fig.~\ref{fig:phaselat}a. Not distinguished on Fig.~\ref{fig:phaselat}a in the hadronic phase are the liquid self-bound ground state of nuclear matter and the gas of free nucleons~\cite{DAgostino05}. 

Predictions for the transition temperature for $\mu_B\sim 0$ are constrained to a relatively narrow range $140 < T_c < 250\,  $MeV, while the critical energy density is predicted to be 5 to 20 times the
normal nuclear energy density, $\epsilon_{0}=0.14$ GeV/fm$^3$. Presumably, the most accurate predictions of the phase transition are given by numerical solutions of the QCD Lagrangian on a lattice~\cite{Creutz}, see Fig.~\ref{fig:phaselat}b~\cite{FodorKatz}. Here, the solid line indicates a first-order phase transition at larger values of $\mu_B\ \gsim\ 360\pm 40$ MeV, with critical endpoint indicated by the small square, followed a smooth crossover for $\mu_B\ \lsim\ 360$ MeV. Interestingly the $\mu_B$ for the critical end point corresponds to $\sqrt{s_{NN}}\simeq 10$ GeV at mid-rapidity, between the maximum and minimum nucleon-nucleon c.m. energies of the AGS and CERN fixed target programs, and is considerably below the value of $\mu_B\simeq 24$ MeV for mid-rapidity at RHIC~\cite{CleymansORWheaton} (see later discussion).

	   A nice feature of the search for the QGP is that it 
requires the integrated use of many disciplines in Physics: High Energy Particle Physics, Nuclear Physics, Relativistic Mechanics, Quantum Statistical Mechanics, and, recently, AdS/CFT string theory~\cite{Policastro,Nastase-BlackHole}. From the point of view of an experimentalist there are two major questions in this field. The first is how to relate the thermodynamical properties (temperature, energy density, entropy, viscosity ...) of the QGP or hot nuclear matter to properties that can be measured in the laboratory. The second question is how the QGP can be detected. 

    One of the
major challenges in this field is to find signatures that are unique to the QGP so that this new state of matter can be distinguished from the ``ordinary physics" of relativistic nuclear collisions.  Another more general challenge is to find effects which are specific to A+A collisions, such as collective or coherent phenomena, in distinction to cases for which  A+A collisions can be considered as merely an incoherent superposition of nucleon-nucleon collisions~\cite{specificity,Weiner05,Alexopoulos02}. 

\section{$J/\Psi$ suppression---the original ``gold-plated" QGP signature}
\begin{figure}[!htb]
\begin{center}
\begin{tabular}{cc}
\includegraphics[width=0.44\linewidth]{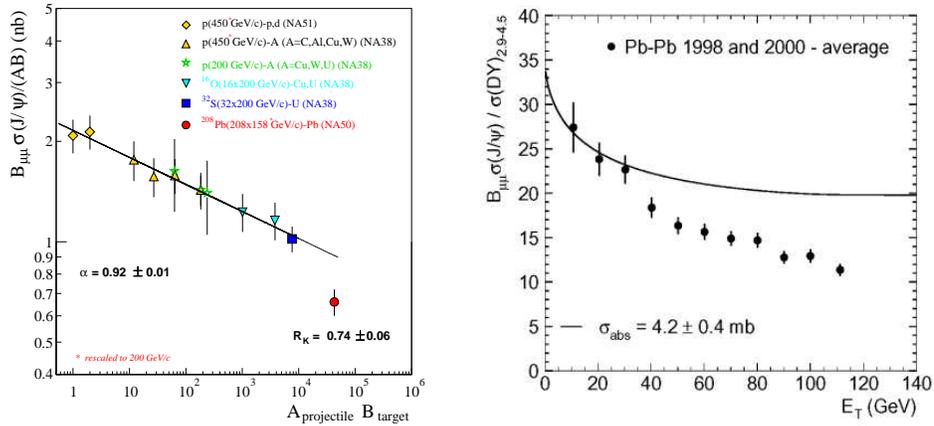}&\hspace*{0.2cm}
\includegraphics[width=0.48\linewidth]{figs/NA50RAAJPSI.epsf}
\end{tabular}
\end{center}\vspace*{-0.25in}
\caption[]{a) (left) Total cross section for $J/\Psi$ production divided by $AB$ in A+B collisions at 158--200$A$ GeV~\cite{NA50PLB450}. b) (right) Cross section for $J/\Psi$ divided by Drell-Yan as a function of centrality measured by $E_T$ for $158A$ GeV Pb+Pb collisions~\cite{NA50EPJC39}. The solid line represents the expected suppression for cold nuclear matter. The plot may be converted to $R_{AA}$ by dividing by 34.   \label{fig:JPsiAB}}
\end{figure}

   Since 1986, the `gold-plated' signature of deconfinement was thought to be $J/\Psi$ suppression. Matsui and Satz~\cite{MatsuiSatz86} proposed that $J/\Psi$ production in A+A collisions will be suppressed by Debye screening of the quark
color charge in the QGP. The $J/\Psi$ is produced when two gluons
interact to produce a $c, \bar c$ pair which then resonates to form the
$J/\Psi$. In the plasma the $c, \bar c$ interaction is screened so that the 
$c, \bar c$ go their separate ways and eventually pick up other quarks at
the periphery to become {\it open charm}. $J/\Psi$ suppression would be quite a spectacular effect since the naive expectation was that $J/\Psi$ production,
due to the relatively large $\sim 1.5$ GeV scale of the charm quark mass, should behave like a pointlike process with production cross section proportional to $A\ (BA)$ for p+A (B+A) minimum bias collisions, and thus
would be enhanced relative to the total interaction cross section,
which increases only as $A^{2/3}$. However, there were problems from the very beginning because unlike Drell-Yan which exhibits an $A^{1.00}$ dependence in p+A collisions, the $J/\Psi$ is suppressed~\cite{E772} by $A^{0.92}$. The suppression continues as $(BA)^{0.92}$ for minimum bias B+A collisions~\cite{NA50PLB450} (Fig.~\ref{fig:JPsiAB}a) at the SPS until the heaviest system, Pb+Pb, where the suppression increases by $\sim 25$\%. However the suppression is much more impressive as function of centrality (Fig.~\ref{fig:JPsiAB}b).
This is the CERN fixed target heavy ion program's main claim to fame. 

\begin{figure}[ht]
\begin{center}
\begin{minipage}[b]{2.4in}
\centerline{Success in HEP}
\centerline{\psfig{file=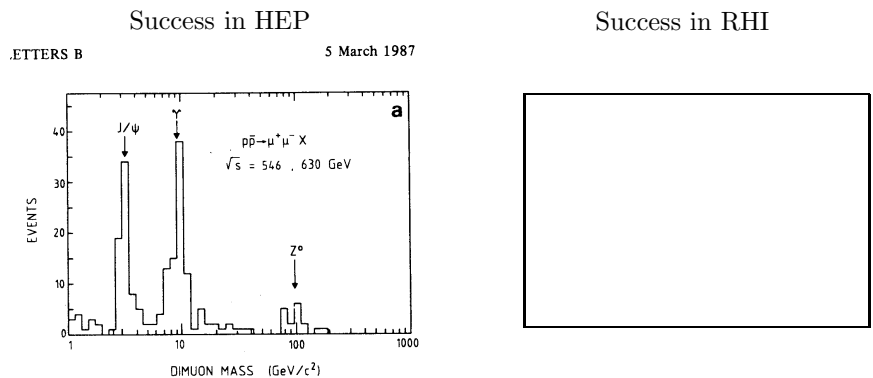,width=2.2in}}
\end{minipage}
\begin{minipage}[b]{2.4in}
\centerline{Success in RHI\hspace*{0.05in}}
\vspace*{0.325in}
\hspace*{0.35in}\begin{picture}(130,88)
\put(0,0){\framebox(130,88){}}
\end{picture}
\vspace*{0.35in}
\end{minipage}
\end{center}\vspace*{-0.12in}
\caption[]{``The road to success": In High Energy Physics (left) a UA1 measurement\cite{UA1} of pairs of muons each with $p_T\geq 3$ GeV/c shows two Nobel prize winning dimuon peaks and one which won the Wolf prize. Success for measuring these peaks in RHI physics is shown schematically on the right. }
\label{fig:success}
\end{figure}

My summary of the different views of dilepton resonances in the High Energy\cite{UA1} and Relativistic Heavy Ion\cite{MatsuiSatz86} Physics communities since the mid 1980's is shown in Fig.~\ref{fig:success}.   

\section{Observables in Relativistic Heavy Ion Collisions}
\begin{figure}[!thb]
\begin{center}
\psfig{file=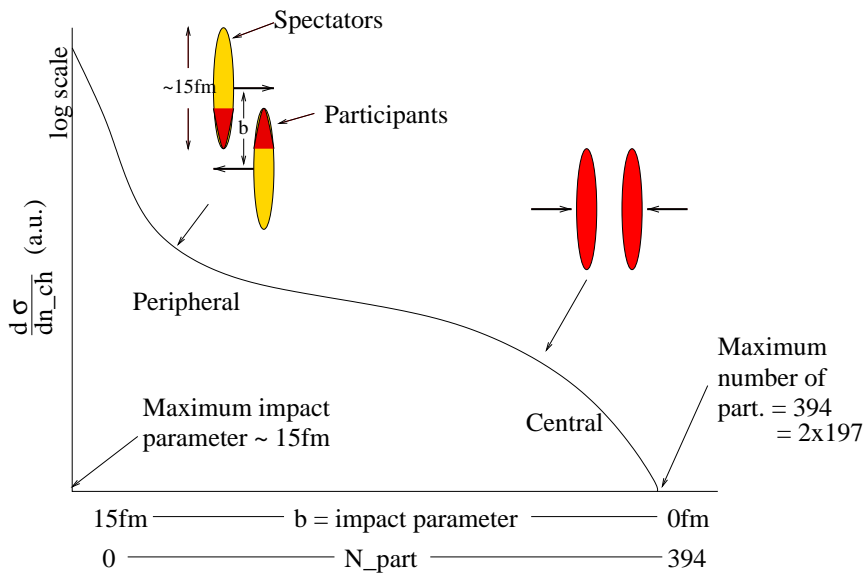,height=0.9\linewidth,angle=-90,bbllx=10, bblly=14, bburx=533, bbury=778,clip=}
\end{center}\vspace*{-0.25in}
\caption[]{a) (left) Schematic of collision of two nuclei with radius $R$ and impact parameter $b$. The curve with the ordinate labeled $d\sigma/d n_{\rm ch}$ represents the relative probability of charged particle  multiplicity $n_{\rm ch}$ which is directly proportional to the number of participating nucleons, $N_{\rm part}$.
\label{fig:nuclcoll}}

\end{figure}

\begin{figure}[!thb]
\begin{center}
\begin{tabular}{cc}
\psfig{file=figs/STARJet+AuAu-g.epsf,width=0.64\linewidth}&\hspace*{-0.025\linewidth}
\psfig{file=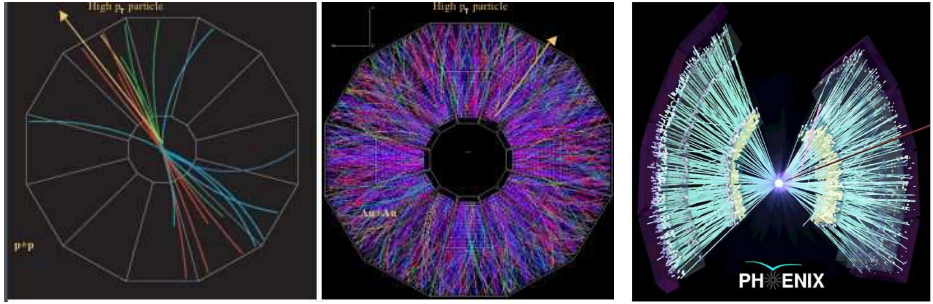,width=0.315\linewidth,height=0.315\linewidth}
\end{tabular}
\end{center}\vspace*{-0.15in}
\caption[]{a) (left) A p-p collision in the STAR detector viewed along the collision axis; b) (center) Au+Au central collision at $\sqrt{s_{NN}}=200$ GeV in the STAR detector;  c) (right) Au+Au central collision at $\sqrt{s_{NN}}=200$ GeV in the PHENIX detector.  
\label{fig:collstar}}

\end{figure}

   A schematic drawing of a collision of two relativistic Au nuclei is shown in Fig.~\ref{fig:nuclcoll}. In the center of mass system of the nucleus-nucleus collision, the two Lorentz-contracted nuclei of radius $R$ approach each other with impact parameter $b$. In the region of overlap, the ``participating" nucleons interact with each other, while in the non-overlap region, the ``spectator" nucleons simply continue on their original trajectories and can be measured in Zero Degree Calorimeters (ZDC), so that the number of participants can be determined. The degree of overlap is called the centrality of the collision, with $b\sim 0$, being the most central and $b\sim 2R$, the most peripheral. The maximum time of overlap is $\tau_O=2R/\gamma\,c$ where $\gamma$ is the Lorentz factor and $c$ is the velocity of light. 
The energy of the inelastic collision is predominantly dissipated by multiple particle production, where $n_{\rm ch}$, the number of charged particles produced, is directly proportional~\cite{PXWP} to the number of participating nucleons ($N_{\rm part}$) as sketched on Fig.~\ref{fig:nuclcoll}. Thus, $n_{\rm ch}$ in central Au+Au collisions is roughly $A$ times larger than in a p-p collision, as shown in actual events from the STAR and PHENIX detectors at RHIC in Fig.~\ref{fig:collstar}.

	          It would appear to be a daunting task to reconstruct all the particles produced in such events. Consequently, it is more common to use single-particle or multi-particle inclusive variables to analyze these reactions. 
     For any observed particle of momentum $\vec{p}$, energy $E$, the momentum can be 
resolved into transverse ($p_T$) and longitudinal ($p_L$) components; and in
many cases the mass ($m$) of the particle can be determined. The longitudinal
momentum is conveniently expressed in terms of the rapidity ($y$):

\begin{equation}
y=\ln\left({E+p_L\over m_T}\right)
\label{eq:1} 
\end{equation}
\begin{equation}
\cosh y=E/m_T \qquad \sinh y=p_L/m_T \qquad dy=dp_L/E
\label{eq:2} 
\end{equation}
where
\begin{equation}
m_T=\sqrt{m^2+p_T^2} \quad {\rm and}\quad 
      E=\sqrt{p_L^2+m_T^2}=\sqrt{p^2+m^2}
\label{eq:3}
\end{equation}
In the limit when ($m\ll E$) the rapidity reduces to the 
pseudorapidity ($\eta$)
\begin{equation}
\eta=-\ln\tan\theta/2
\label{eq:4} \end{equation}
\begin{equation}
\cosh\eta=\csc\theta \qquad \sinh\eta=\cot\theta
\label{eq:5}
\end{equation}
where $\theta$ is the polar angle of emission.
The rapidity variable has the useful property that it 
is additive under a Lorentz transformation.

\section{The RHIC facility}

\begin{figure}[!th]
\centerline{\psfig{file=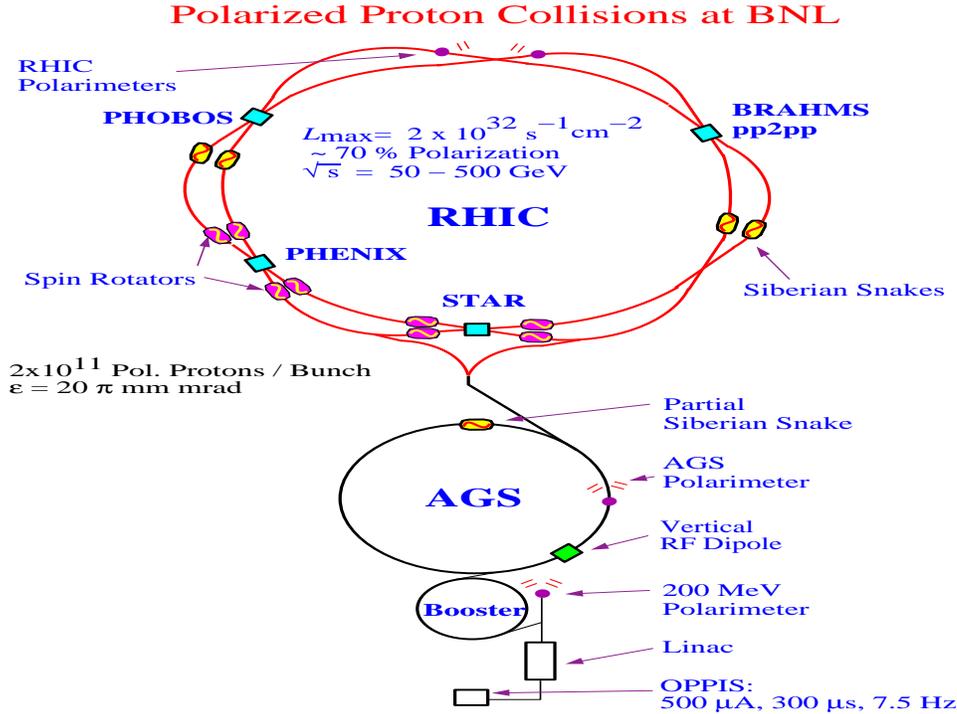,height=0.75\linewidth,width=\linewidth}}
\vspace*{8pt}
\caption[]{Schematic of RHIC accelerator with emphasis on equipment for polarized proton acceleration and storage. For Heavy Ions, injection to the booster is via transfer line from the Tandem Van de Graaf.}
\label{fig:RHIC}
\end{figure}

   The Relativistic Heavy Ion Collider (RHIC)~\cite{RHICNIM} at Brookhaven National Laboratory consists of two independent superconducting acceleration/storage rings of 3.8km circumference which cross at six interaction regions where experiments can be performed (see Fig.~\ref{fig:RHIC}). At present, RHIC can collide any nucleus  from proton to Au with any other nucleus, preferably at equal velocity $\beta=v/c$ ($\gamma=1/\sqrt{1-\beta^2}$), with an achieved luminosity of $4\times 10^{26}$ cm$^{-2}$ sec$^{-1}$ for Au+Au collisions at a beam energy of 100 GeV/nucleon which corresponds to a nucleon-nucleon collision c.m. energy $\sqrt{s_{NN}}=200$ GeV. For p-p collisions, the beams are always polarized, with a design polarization of 70\% per beam, with 65\% achieved at present with a maximum luminosity of $2\times 10^{31}$ cm$^{-2}$ sec$^{-1}$ at $\sqrt{s}=200$ GeV. The maximum c.m. energy for p-p collisions is $\sqrt{s}=500$ GeV, with a luminosity of $> 10^{32}$ cm$^{-2}$ sec$^{-1}$. In general, the luminosity in AB collisions is proportional to $\gamma^2/(AB)$ from the injection energy of $\sim 24\ Z/A$ GeV to the maximum energy of $250\ Z/A$ GeV and is proportional to $\gamma^{3}-\gamma^{4}$ for proposed deceleration below the injection energy to allow Au+Au collisions with $\sqrt{s_{NN}}$ as low as $\gsim\ 5$ GeV to access the region of the possible critical point (Fig.~\ref{fig:phaselat}b). In the future, electron cooling will increase the luminosity by a factor $\geq 20$ and a new Electron Beam Ion Source (EBIS) to replace the Tandem Van de Graaf and transfer line will allow nuclei up to Uranium~\cite{RHICNIM}. A summary of the different species, energies and data collected in the first six years of RHIC operation is given in Table~\ref{tab:RHIChistory}.
\begin{table}[h]
\caption[]{Summary of RHIC runs: Date, conditions, Integrated Luminosity, number of scanned interactions ($N_{\rm tot}$ int.), equivalent p-p Luminosity for Hard-Scattering, Data Archived.\label{tab:RHIChistory}}  
\begin{tabular} {|r|rrrrrrr|}
\hline
Run & Year & Species & $\sqrt{s_{NN}}$ & $\int L dt\quad$ & N$_{\rm tot}$ & p-p Equiv. & Data \\
    &      &         &     GeV     &                  &  int.         &   $AB\times\int L dt$ & Size \\
\hline
01 & 2000 & Au+Au & 130  & 1 $\mu$b$^{-1}$ & 10M & 0.04 pb$^{-1}$ & 3 TB\\
02 & 2001--2 & Au+Au & 200  & 24 $\mu$b$^{-1}$ & 170M & 1.0 pb$^{-1}$ & 10 TB\\
   &           & p+p   & 200  & 0.15 pb$^{-1}$ &   3.7G & 0.15 pb$^{-1}$ & 20 TB\\
03 & 2002--3 & d+Au  & 200  & 2.74 nb$^{-1}$ & 5.5G  & 1.1 pb$^{-1}$ &   46 TB\\
   &         & p+p   & 200  & 0.35 pb$^{-1}$ & 6.6G  & 0.35 pb$^{-1}$ & 35 TB\\
04 & 2003--4 & Au+Au & 200  & 241 $\mu$b$^{-1}$ & 1.5G & 10.0 pb$^{-1}$ & 270 TB\\  
   &         & Au+Au &  62  & 9 $\mu$b$^{-1}$ & 58M  & 0.36 pb$^{-1}$ &  10 TB\\
05 & 2004--5 & Cu+Cu & 200  & 3 nb$^{-1}$ & 8.6G & 11.9 pb$^{-1}$ & 173 TB\\
   &         & Cu+Cu &  62  & 0.19 nb$^{-1}$ & 0.6G & 0.8 pb$^{-1}$ & 48 TB\\
   &         & Cu+Cu & 22.5 & 2.7 $\mu$b$^{-1}$ & 9M & 0.01 pb$^{-1}$ & 1 TB \\
   &         & p+p   & 200  & 3.8 pb$^{-1}$ & 85B & 3.8 pb$^{-1}$ & 270 TB\\
06 & 2006 & p+p  & 200   & 10.7 pb$^{-1}$ & 230B & 10.7 pb$^{-1}$ & 310 TB\\
             &      & p+p  &  62   & 0.1 pb$^{-1}$ & 28B & 0.1 pb$^{-1}$ &  25 TB\\ 
\hline
      \end{tabular}
      \end{table}
      
\subsection{The Experiments}

   There are two major experiments at the RHIC heavy ion program, STAR~\cite{STWP} and PHENIX~\cite{PXWP}, and two smaller experiments. The two small experiments, which have now completed their programs, are PHOBOS~\cite{PHWP}, which emphasizes charged particle detection over the full phase space using Si detectors, and BRAHMS~\cite{BRWP}, with two small-aperture, high precision moveable spectrometers, one at mid-rapidity and one capable of moving as far forward as $2^{\circ}$ ($|\eta|>4$). 
   
STAR, which emphasizes hadron physics, is most like a conventional hadron-collider detector, a TPC covering the 
full azimuth over \mbox{$\pm 1$} unit of pseudorapidity, for the 
purpose of charged particle tracking in a magnetic field of 0.5 Tesla. 
The TPC is  surrounded by a 
system of Time of Flight counters, for particle identification, and a 
moderate resolution ($15\% /\sqrt{E}$) 
electromagnetic calorimeter, for measuring $\pi^0$ production and 
charged-neutral energy correlations. The detector is completed by a 
Silicon Drift 
Vertex Tracker, for measurements of Hyperons, and possible TPC's external to 
the magnet, for tracking at small angles $2.0\leq |\eta| \leq 4.5$. 

      PHENIX, a very high granularity, high resolution detector
for leptons and photons emerging from the Quark Gluon Plasma (QGP), emphasizes 
the ability to run at the highest luminosities with very selective triggers to 
find rare events, particularly $J/\Psi$, open charm and direct photon production.  PHENIX has a highly instrumented
electron, photon 
and charged hadron spectrometer, in the central region $|\eta|\leq 0.35$, with 
full azimuth di-muon measurement in two endcaps, $1.15\leq |\eta| \leq 2.35$. 
The 
electron/photon central spectrometer emphasizes electron identification 
at the trigger level, with RICH, TRD and EM calorimetry. The EM calorimeter, 
with energy resolution $\sigma_E/E=7\%/\sqrt{E({\rm GeV})}$,   
also serves as an excellent photon and $\pi^0$ trigger because of its 
5 by 5 cm segmentation at 5.1 m.  
The central spectrometer consists of 
two arms, each subtending $90^\circ$ in azimuth ($\Phi$) and $\pm 0.35$ units 
in pseudorapidity ($\eta$). The total coverage is 1/2 of the azimuth---however, 
the two arms are not back-to-back: the gap between the edges of the two 
$90^\circ$ arms is $67.5^\circ$ on one side and hence $112.5^\circ$ on the 
other.  The charged particle momentum 
resolution is 1\% at 5 GeV/c, and charged hadron identification is 
provided by TOF(100ps) for 1/3 of the azimuth of one arm.

\begin{figure}[!t]
\begin{tabular}{cc}
\centering\psfig{file=figs/ERT-YGoto.epsf,width=0.52\linewidth}
\hspace{0.014in}
\centering\psfig{file=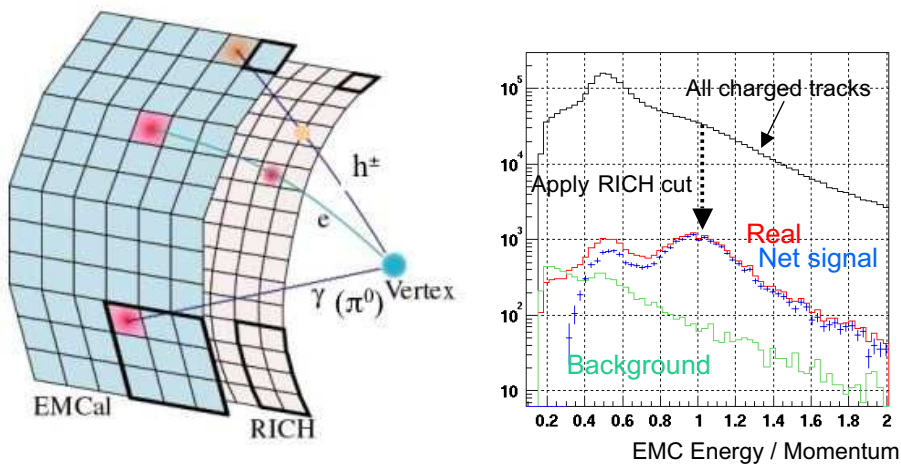,width=0.44\linewidth}
\end{tabular}
\caption[]{a (left) Schematic of $\pi^{\pm}$, $e^{\pm}$ and $\gamma$ in PHENIX, with ElectroMagnetic Calorimter (EMCal) and Ring Imaging Cherenkov Counter (RICH). b) (right) Energy/momentum for all charged particles detected in the EMcal with and without a RICH signal.}
\label{fig:pxworks}
\end{figure}

	Rather than show a diagram of PHENIX, which is a very non-conventional collider detector~\cite{egseePT}, the principles of the design of PHENIX will be illustrated (see Fig.~\ref{fig:pxworks}). 
In order to detect electrons in hadron collisions, where the typical ratio of $e^{\pm}/\pi^{\pm}$ is known to be of order of $10^{-4}$ for prompt electrons (from charm) at $23.5 \leq \sqrt{s}\leq 62.4$ GeV~\cite{CCRS} (ISR energies), one must plan on a charged pion rejection of $ > 10^5$. PHENIX decided to use a Ring Imaging Cerenkov counter (RICH) in combination with an Electromagnetic Calorimeter (EMCal) to achieve this rejection. There is also the issue of the huge background of $e^{\pm}$ from internal and external conversion of the photons from $\pi^0\rightarrow \gamma+\gamma$ decay or from direct photon production, which must be measured and understood to high precision. The axial magnetic field in PHENIX is designed with the possibility of zero magnetic field on the axis---so that the lower energy member of an asymmetric $e^+ e^-$ conversion pair does not curl up and get lost---and with a minimum of material in the aperture (0.4\% of a radiation length) to avoid external conversions. The EMCal is crucial for electron identification and hadron rejection.  
	The EMCal measures the energy of $\gamma$ and $e^{\pm}$ and reconstructs $\pi^0$ from 2 photons. The high granularity of the individual EMCal towers is $(\delta\eta \times \delta\phi)$ $\sim (0.01\times 0.01)$, which allows the two photons from a $\pi^0$ to be resolved from a single photon cluster for values of $p_T$  up to $\gsim\ 30$ GeV/c. It measures a decent time of flight (TOF), 0.3 nanoseconds over 5 meters, allowing photon and charged particle identification. The high precision TOF over part of the aperture allows improved charged hadron identification. Electrons are identified by a count in the RICH and matching Energy and momentum ($E/p$), where the momentum is measured by track chambers in a magnetic field. Charged hadrons deposit only minimum ionization in the EMCal ($\sim 0.3$ GeV), or higher if they interact, and don't count in the RICH ($\pi^{\pm}$ threshold 4.7 GeV/c). Thus, requiring a RICH signal rejects all charged hadrons with $p < 4.7$ GeV/c, leaving only $e^{\pm}$ as indicated by the $E/p=1$ peak in Fig.~\ref{fig:pxworks}-right. $\pi^{\pm}$ can be identified  above 4.7 GeV/c by a RICH signal together with an EMCal hit with minimum ionization or greater. 
	It is amusing to realize that once you decide to measure electrons, you must make an excellent $\pi^0$ measurement to understand the background, and this implies a detector which can measure and identify almost all particles, as exemplified by the PHENIX central spectrometer.

\section{Inclusive and semi-inclusive particle production}

A single particle ``inclusive'' reaction involves the measurement of just one
particle coming out of a reaction,  
\[ a + b \rightarrow c +\mbox{\rm anything} \;\;\; .\]
The terminology~\cite{FeynmanScaling} comes from the fact that all final states with the particle 
$c$ are summed over, or {\em included}. A ``semi-inclusive'' 
reaction\cite{KNO} refers to the measurement of all events of a given  
topology or class, e.g. 
\[ a + b \rightarrow n_1 \mbox{\rm \ particles in class 1}  
+\mbox{\rm anything} \;\;\; ,\]
where ``centrality'' is the most common class in relativistic heavy ion collisions.

\begin{figure}[htb]
\begin{center}
\begin{tabular}{cc}
\psfig{file=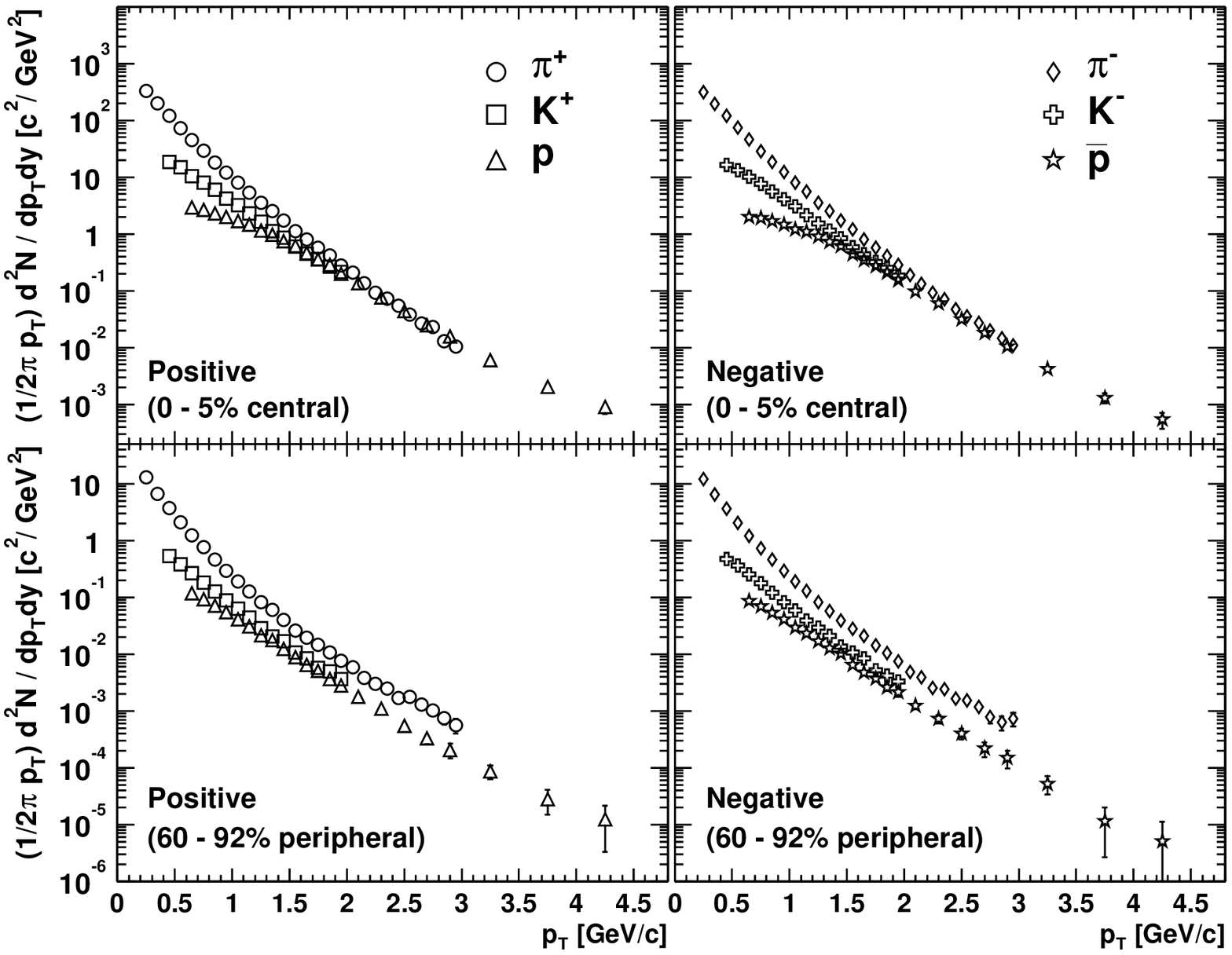,width=0.65\linewidth}
&\hspace*{-0.35in}
\psfig{file=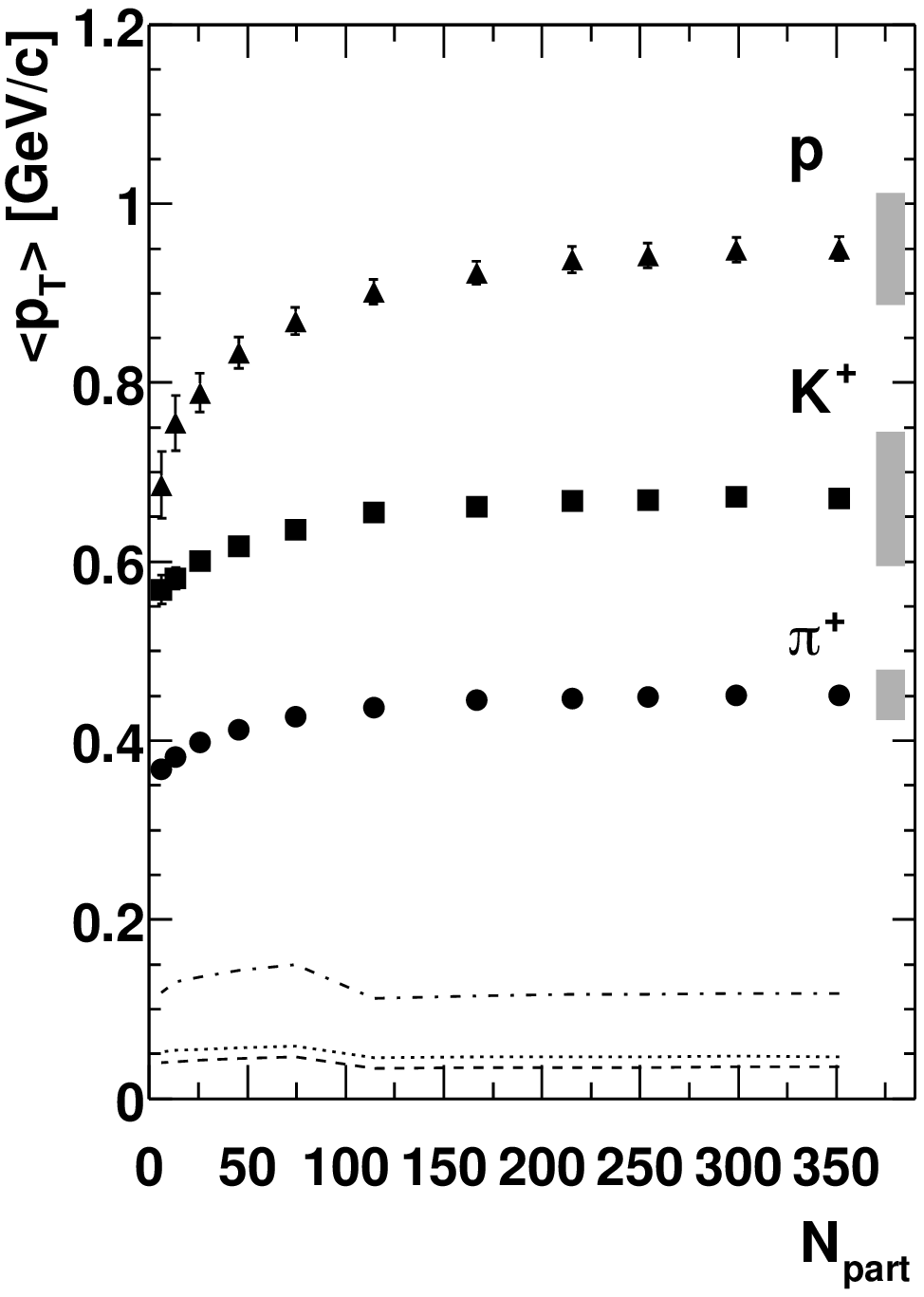,width=0.33\linewidth}
\end{tabular}
\end{center}\vspace*{-0.25in}
\caption[]{a) (left) Semi-inclusive invariant $p_T$ spectra for $\pi^{\pm}$, $K^{\pm}$, $p^{\pm}$ in Au+Au collisions at $\sqrt{s_{NN}}=200$ GeV~\cite{PXWP,PXPRC69pid}.  
b) (right) $\mean{p_T}$ of positive particles as a function of centrality ($N_{\rm part}$) from the same data.
\label{fig:PXpTspectra} }
\end{figure}

	 Measurements are presented in terms of 
the (Lorentz) invariant single particle inclusive differential cross section (or Yield per event in the class if semi-inclusive):
\begin{equation}
{Ed^3\sigma\over dp^3}={d^3\sigma\over p_T dp_T dy d\phi}=
{1\over 2\pi}\, {\bf f}(p_T,y) \quad ,
\label{eq:siginv} \end{equation} 
where $y$ is the rapidity, $p_T$ is the transverse momentum, and $\phi$ is the azimuth of the particle (see Fig.~\ref{fig:PXpTspectra}). 
The average transverse momentum, $\mean{p_T}$, or the mean transverse kinetic 
energy, $\mean{m_T}- m$, or the asymptotic slope are taken as measures of the 
temperature, $T$, of the reaction.  

	It is important to be aware that the integral of the single particle inclusive cross section over all the variables is not equal to $\sigma_I$ the interaction cross section, but 
rather is equal to the mean multiplicity times the interaction cross section: 
$\mean{n} \times\, \sigma_I$. Hence the mean multiplicity per interaction is 
\begin{equation}
\mean{n} ={1\over \sigma_I}\, \displaystyle\int {d\phi \over {2\pi}} dy\,dp_T\,\, p_T\,\, {\bf f}(p_T,y)= {1\over \sigma_I}\int dy\,{{d\sigma}\over {dy}} 
=\int dy\, \rho(y) \;\;\; , 
\label{eq:P4} \end{equation}
where the terminology for the multiplicity density in rapidity is $(1/\sigma_I)\,d\sigma/dy= \rho(y)=dn/dy$ for identified particles ($m$ known), $dn/d\eta$ for non-identified particles ($m$ unknown, assumed massless). The total charged particle multiplicity is taken as a measure of the total entropy, $s$ and $dn/d\eta$ is taken as a measure of the entropy density in restricted intervals of rapidity.  

\subsection{ The rapidity density and the rapidity plateau.} 
At RHIC, the reference frame of the detectors is designed to be the nucleon-nucleon c.m. frame in which the two nuclei approach each other with the same $\gamma$. The nucleon-nucleon c.m. energy is denoted $\sqrt{s_{NN}}$, and the total c.m. energy is  $\sqrt{s}=A\cdot\sqrt{s_{NN}}$ for symmetric A+A collisions. The colliding nucleons approach each other with energy $\sqrt{s_{NN}}/2$ and equal and opposite momenta. The rapidity of the nucleon-nucleon center of mass (mid-rapidity) is $y^{\rm cm}=y_{NN}=0$, and the projectile and target nucleons are at equal and opposite rapidities: 
    \begin{equation}
y^{\rm proj}=-y^{\rm target}=\cosh^{-1}\,{\sqrt{s_{NN}} \over {2 m_N}}=y^{\rm beam}/2\qquad ,  
\label{eq:syk9}
\end{equation} 
where $m_N=$931~MeV is the mass of nucleon bound in a nucleus. \footnote{The Laboratory reference frame used in fixed target experiments is shifted by $y^{\rm beam}/2$ from the c.m. system such that $y_{L}^{\rm target}=0$, $y_L^{\rm proj}=y^{\rm beam}$, $y_L^{\rm cm}=y^{\rm beam}/2$.}

\begin{figure}[!hbt]
\begin{center}
\hspace*{0.23\linewidth}\psfig{file=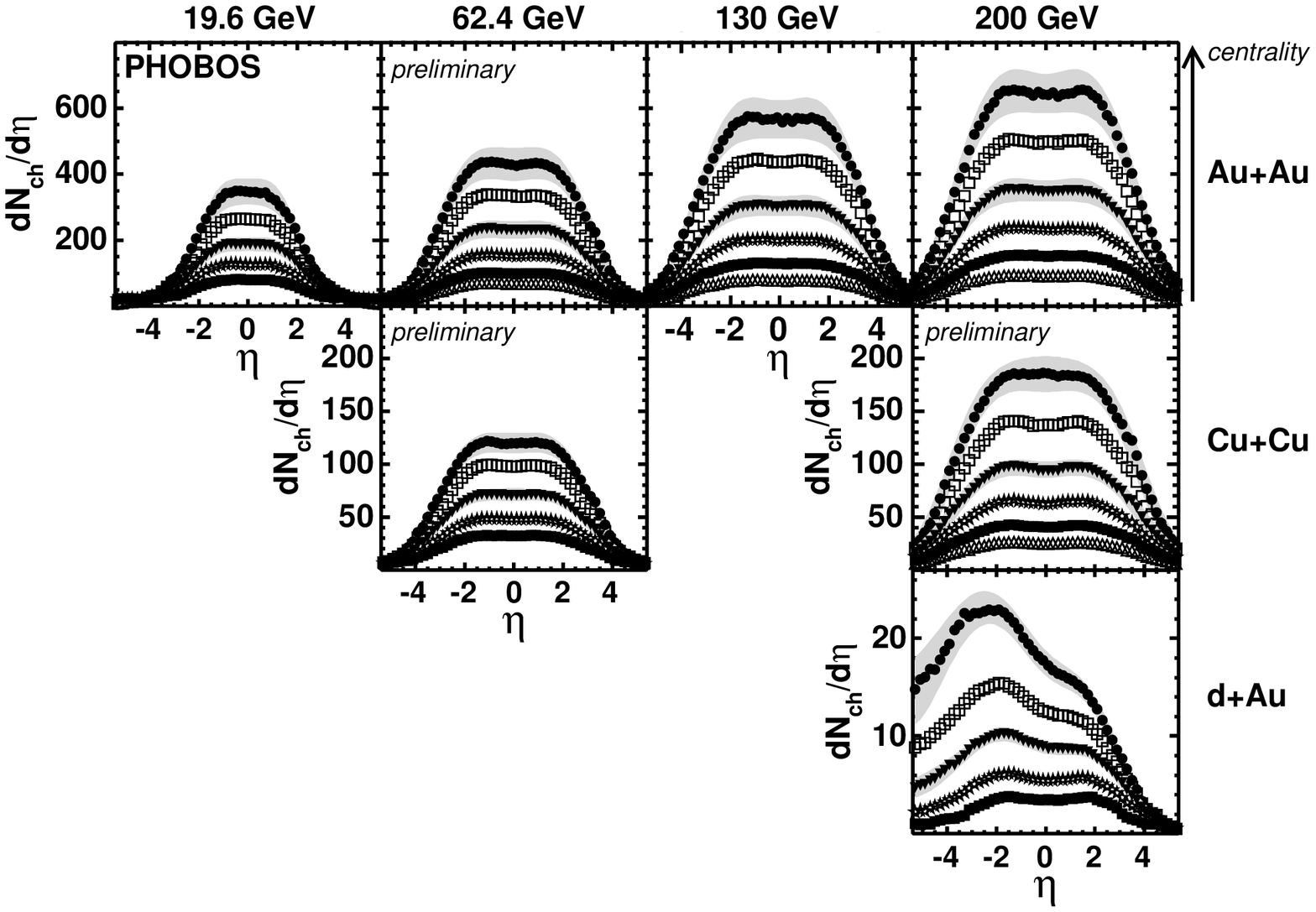,width=0.77\linewidth}
\end{center}
\vspace*{-0.85in}
\psfig{file=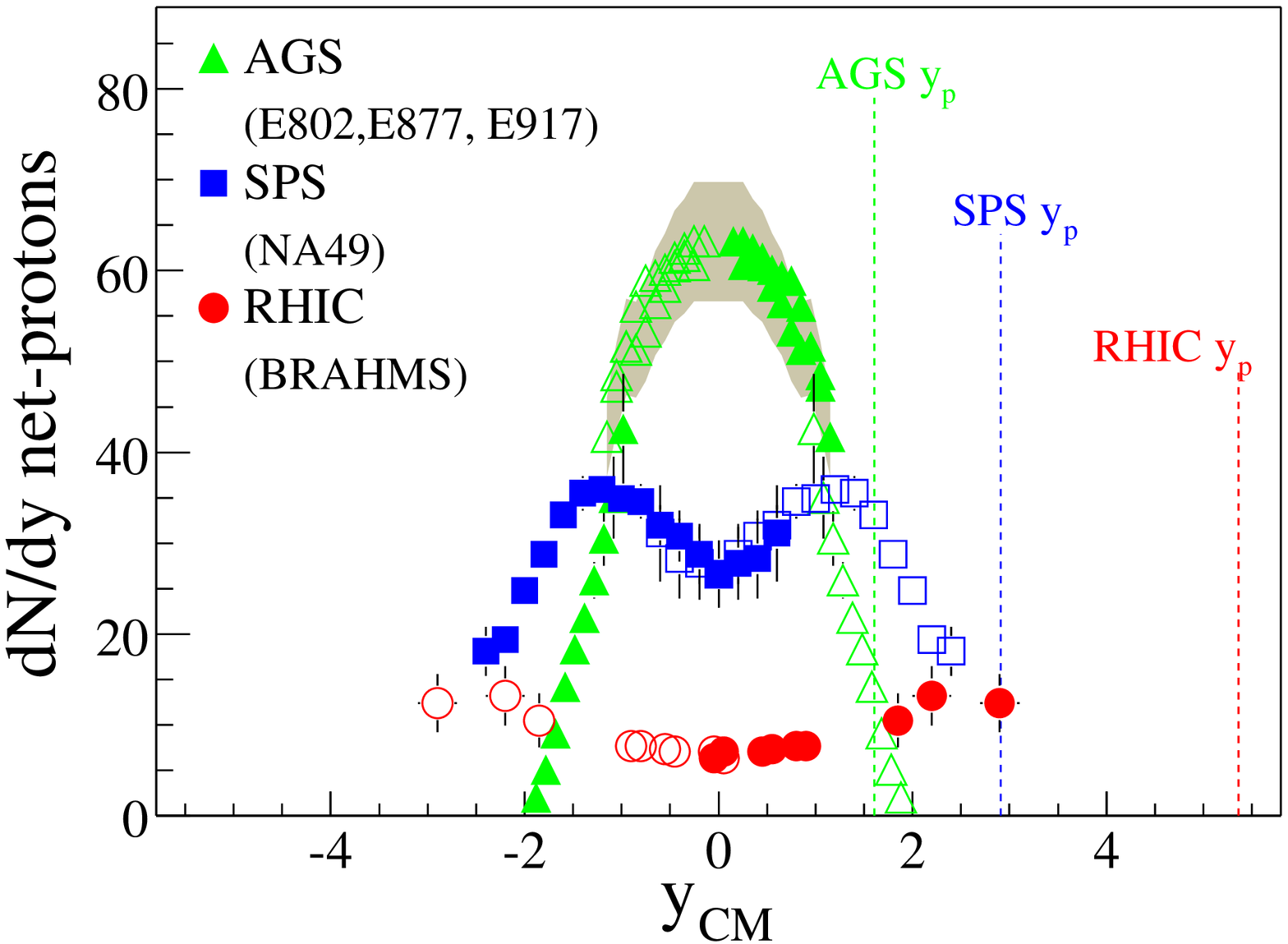, width=0.5\linewidth}
\caption[]{a) (top right) $dn/d\eta$ for A+A and d+Au collisions at RHIC as a function of $\sqrt{s_{NN}}$~\cite{RolandQM05}. b) (bottom left) $dn/dy|_{p} - dn/dy|_{\bar{p}}$ at AGS, SPS and RHIC, $y^{\rm proj}=1.6, 2.9, 5.4$~\cite{BrahmsPRL93}. 
\label{fig:PHOBOSQM05}}

\end{figure}

	The shape and evolution with $\sqrt{s}$ of the charged particle 
density in rapidity, $dn/dy$, in p-p, p+A and A+A collisions all follow a similar trend 
(Fig~\ref{fig:PHOBOSQM05}a)~\cite{RolandQM05} and provide a graphic description of high energy 
collisions. Regions of nuclear fragmentation take up the first 1-2 units around the projectile and target rapidity and if the center-of-mass energy is sufficiently high, a
central plateau is exhibited.   
The distributions increase in width with increasing $\sqrt{s_{NN}}$ but by a smaller amount than the increase in $y^{\rm beam}$ and show a small decrease in width with increasing centrality. $dn/d\eta$ increases with increasing centrality, $\sqrt{s_{NN}}$ and A in A+A collisions. In the asymmetric d+Au collision, $dn/d\eta$ in the target rapidity region is larger than in the projectile region, but not by much, only about ~50\%. Also the nuclear transparency is evident, there is no reduction of particles at the projectile rapidity with increasing centrality. 

	Subtleties of the distributions in A+A collisions become apparent when identified particles are used~\cite{BrahmsPRL93}. In Fig.~\ref{fig:PHOBOSQM05}b, the difference of $dn/dy$ for protons and anti-protons, i.e net-protons is shown as a function of c.m. energy, 
$\sqrt{s_{NN}}=$ 5 (AGS, Au+Au), 17 (SPS, Pb+Pb), 200 (RHIC, Au+Au) GeV, $y^{\rm proj}=$1.6, 2.9, 5.4. As $\sqrt{s_{NN}}$ is reduced,  
stopping 
of the participating nucleons 
is indicated by the nuclear fragmentation peak moving from the fragmentation region (not visible for RHIC) to mid-rapidity.

	\subsection{Centrality Measurement, Fluctuations, Bjorken Energy Density}
	\begin{figure}[hbt]
\centering\psfig{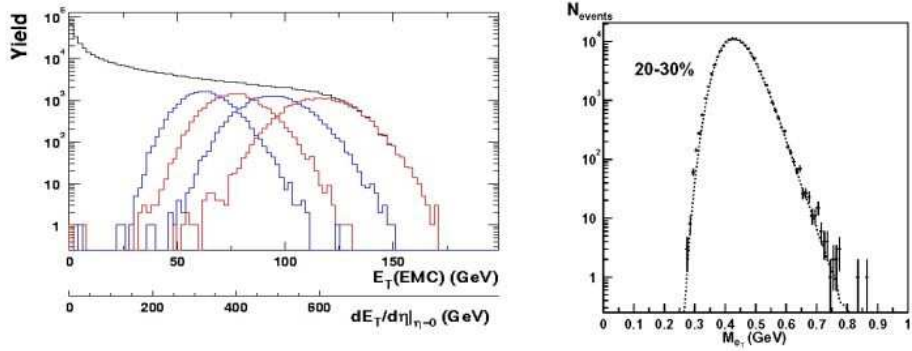}
\vspace*{-0.12in}
\caption[]{a) (left) PHENIX~\cite{PXppg002} $E_T$ distribution in Au+Au collisions at $\sqrt{s_{NN}}=200$ GeV showing centrality selections of 0-5\%, 5-10\%, 10-15\%, 15-20\%. b) (right) Event-by-event average $E_T$/cluster in the EMCal for 20-30\%  centrality~\cite{PXPRC66} (data points), random baseline (dotted line).}
\label{fig:ETfluc}
\end{figure}
           Another variable, closely related to multiplicity,  is the
transverse energy density in rapidity or $dE_T/dy \sim \mean{p_T}\times dn/dy$, usually measured in calorimeters by summing over all particles on an event in a fixed but relatively large solid angle~\cite{MJTErice03}: $E_T=\sum_i E_i \sin\theta_i$.  
$dE_T/dy$ is thought to be related to the co-moving energy density in
a longitudinal 
expansion~\cite{BjorkenPRD27,PXWP}, and taken by all experiments as a measure of 
the energy density in space $\epsilon$:
\begin{equation}
\epsilon_{Bj}={d\mean{E_T}\over dy} {1\over \tau_F\pi R^2}
 \label{eq:eBj}
 \end{equation}
where $\tau_F$, the formation time, is usually taken as 1 fm/c,
$\pi R^2$ is the effective area of the collision, and $d\mean{E_T}/dy$ is the 
co-moving energy density.

	Besides the Bjorken energy density, the importance of $E_T$ distributions in RHI collisions (see Fig.~\ref{fig:ETfluc}a) is that they are sensitive primarily to the nuclear geometry of the reaction, and hence can be used to measure the centrality of individual interactions on an event-by-event basis. However, in PHENIX, centrality is determined far from mid-rapidity to avoid possibly biasing any of other the mid-rapidity measurements. The centrality is determined by the upper percentiles of the charged multiplicity distribution in Beam Beam Counters (BBC) which cover the region $3.0<|\eta|<3.9$ in coincidence with Zero Degree Calorimeters located at $|\eta|>6$. The shift of the $E_T$ distributions to higher values with increasing centrality (smaller upper-percentiles) is nicely illustrated in Fig.~\ref{fig:ETfluc}a. For 0-5\% centrality the Bjorken energy density is 5.5 GeV/fm$^3$, well above the value of 1 GeV/fm$^3$ nominally envisioned for QGP formation. 
	
Fluctuations can also be studied. For instance, is the shape of the upper edge of the centrality selected $E_T$ distributions in Fig.~\ref{fig:ETfluc}a random, or is it evidence of non-random fluctuations? The event-by-event distribution $M_{e_T}$ of the average $E_T$ per cluster, which is closely related to $E_T$ and also to $M_{p_T}$, the event-by-event average $p_T$ of charged particles, is shown in Fig.~\ref{fig:ETfluc}b~\cite{PXPRC66}. The non-random fluctuation is measured by comparison to a random baseline from mixed events.  $M_{p_T}$ is defined: 
 \begin{equation}
M_{p_T}=\overline{p_T}={1\over n} \sum_{i=1}^n p_{T_i}={1\over 
n} E_{Tc} \qquad ,\label{eq:defMpT}
\end{equation}
where $E_{Tc}$ is the analog of $E_T$ for charged particles. 
PHENIX~\cite{PXppg027} has shown that the non-random $M_{p_T}$ fluctuations are of the order of 1\%.  This places severely small limits on the critical fluctuations expected for a first order phase transition but is consistent with the lattice QCD prediction of a smooth crossover at RHIC energies (Fig.~\ref{fig:phaselat}b).
  
\subsection{Particle abundances, thermal/chemical equilibrium} 
\begin{figure}[!thb]
\begin{center}
\begin{tabular}{cc}
\psfig{file=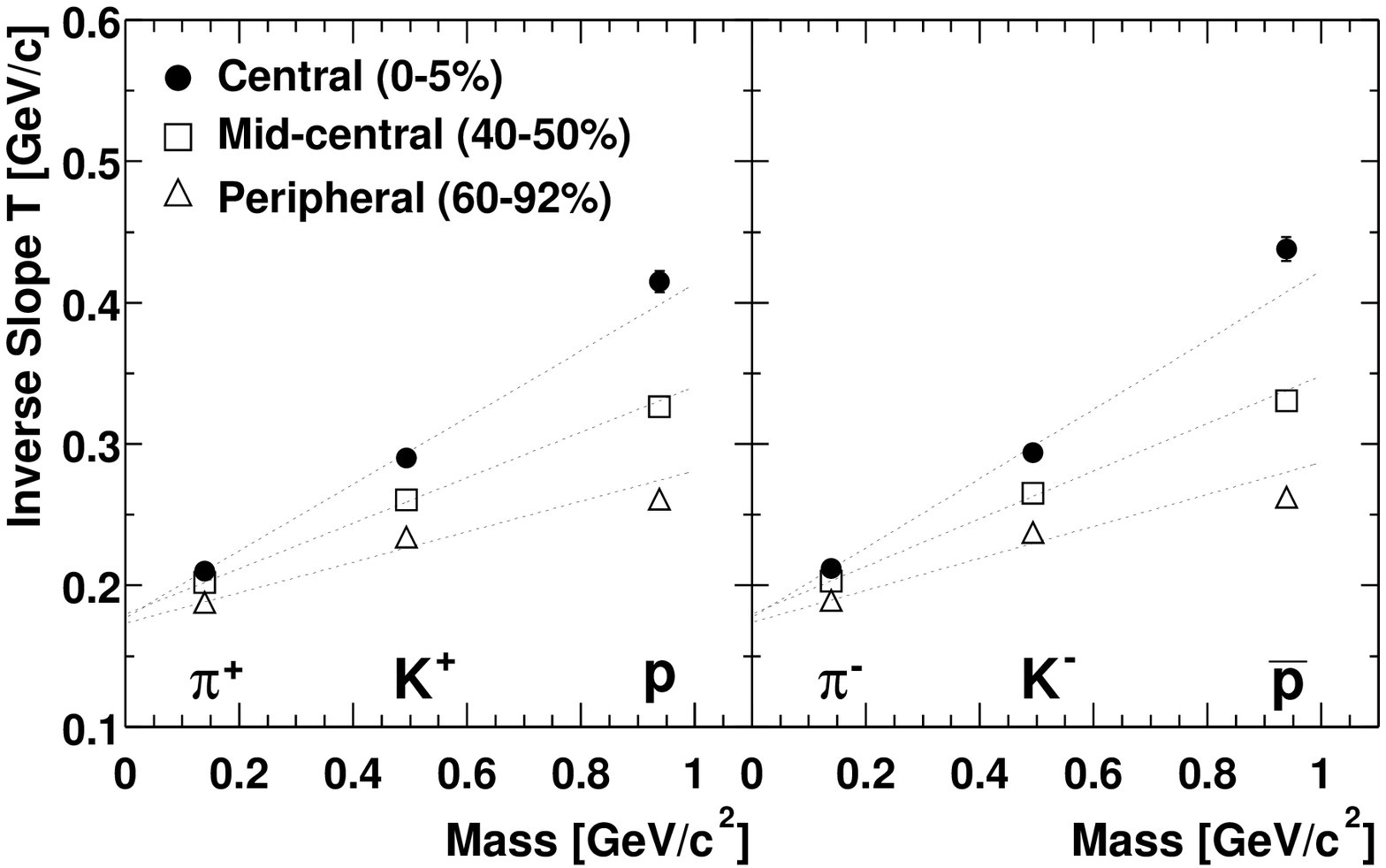,width=0.42\linewidth,height=0.46\linewidth}&\hspace*{-0.06\linewidth}
\psfig{file=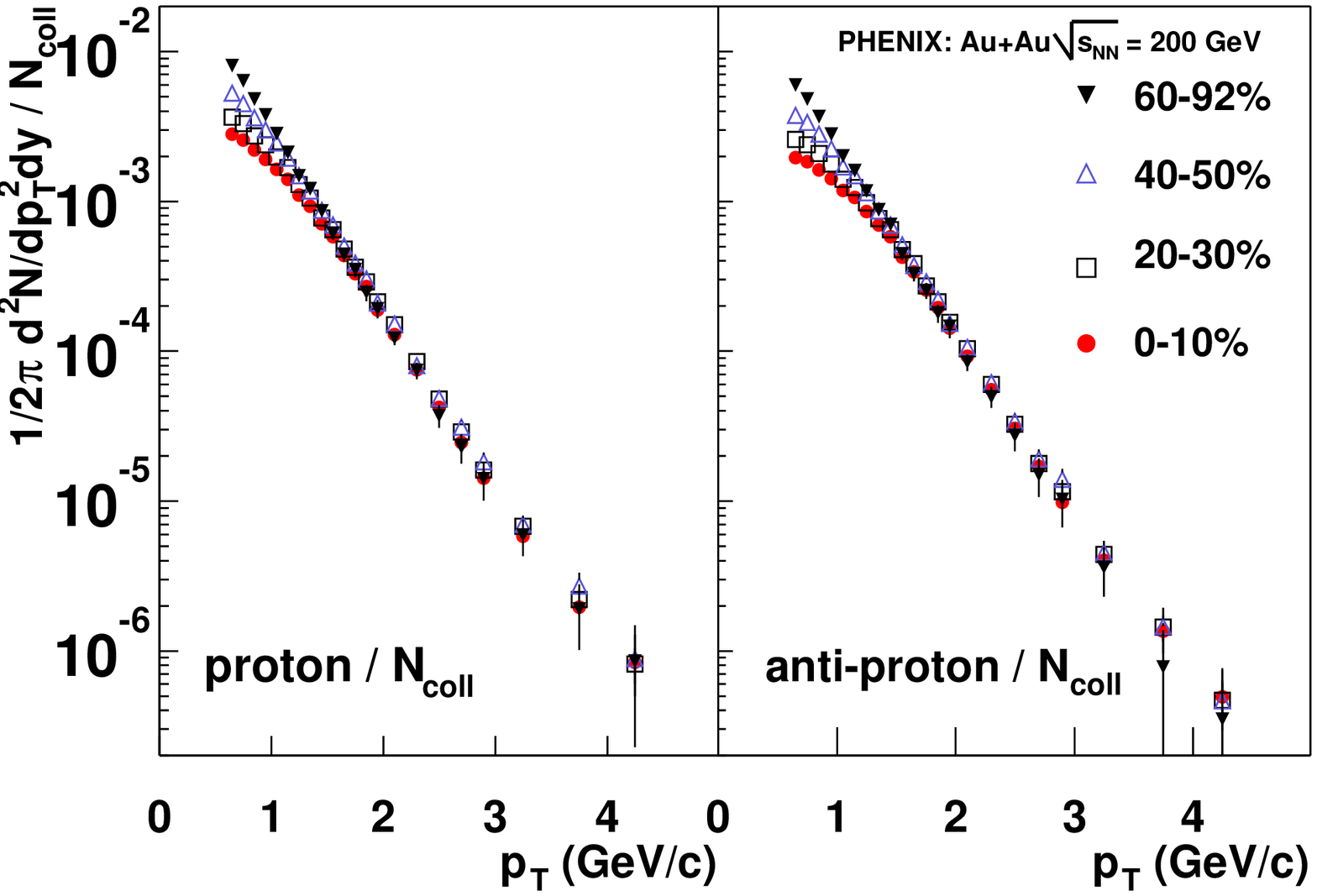,width=0.58\linewidth}\end{tabular}
\end{center}\vspace*{-0.15in}
\caption[]{a) (left) Inverse slope of the $m_T$ distribution~\cite{PXPRC69pid}  (colloquially called $T$) for $\pi^{\pm}$, $K^{\pm}$, $p$, $\bar{p}$ in Au+Au collisions for three centralities at $\sqrt{s_{NN}}=200$ GeV. b) (right) Inviarant yield of $p$ and $\bar{p}$ as a function of centrality scaled (by the number of binary collisions $N_{\rm  coll}$) to lie on top of each other for $p_T\gsim 2$ GeV/c~\cite{PXscalingPRL91}        \label{fig:radialflow}}
\end{figure}

    As shown in Fig.~\ref{fig:PXpTspectra}b, the $\mean{p_T}$ of $\pi^{\pm}$, $K^{\pm}$, $p$, $\bar{p}$, at RHIC, increases smoothly from peripheral to  central Au+Au collisions, and as in pp collisions increases with increasing mass as would be expected for a thermal distribution (Eq.~\ref{eq:boltz}):  
    \begin{equation}
{{d^2\sigma} \over {dp_L p_T dp_T}}={{d^2\sigma} \over {dp_L m_T dm_T}}={1\over {e^{E/T} \pm 1}}\sim e^{-E/T} \qquad .
\label{eq:boltz}
\end{equation}
Since $E=m_T \cosh y$, a signal of thermal production is that the $p_T$ and mass dependence of the cross section are not independent
but depend only on the transverse mass $m_T$. Thus, the inverse-slope of the $m_T$ distribution at mid-rapidity should represent the temperature. However,  
    the $m_T$-scaling property of thermal distributions is not exactly exhibited by the data (see Fig.~\ref{fig:radialflow}a). 
The inverse slopes increase linearly with rest mass, by a larger amount with increasing centrality, which is evidence for collective motion (`radial flow') and is seen in RHI collisions at AGS~\cite{AkibaQM96}, SPS~\cite{NA44PRL78} and RHIC~\cite{BRWP,PHWP,STWP,PXWP} energies. Particles (or partons) which travel with a transverse flow velocity $\beta_{T}$ acquire kinetic energy in addition to the thermal energy so that the inverse slope should increase linearly with the rest mass, $T\rightarrow T_0 +\gamma_T\ m_0$, as illustrated by the lines on Fig.~\ref{fig:radialflow}a. The effect is primarily at low $p_T$ where the slope flattens with increasing centrality as illustrated in Fig.~\ref{fig:radialflow}b. 

\begin{figure}[!thb]
\begin{center}
\begin{tabular}{cc}

\psfig{file=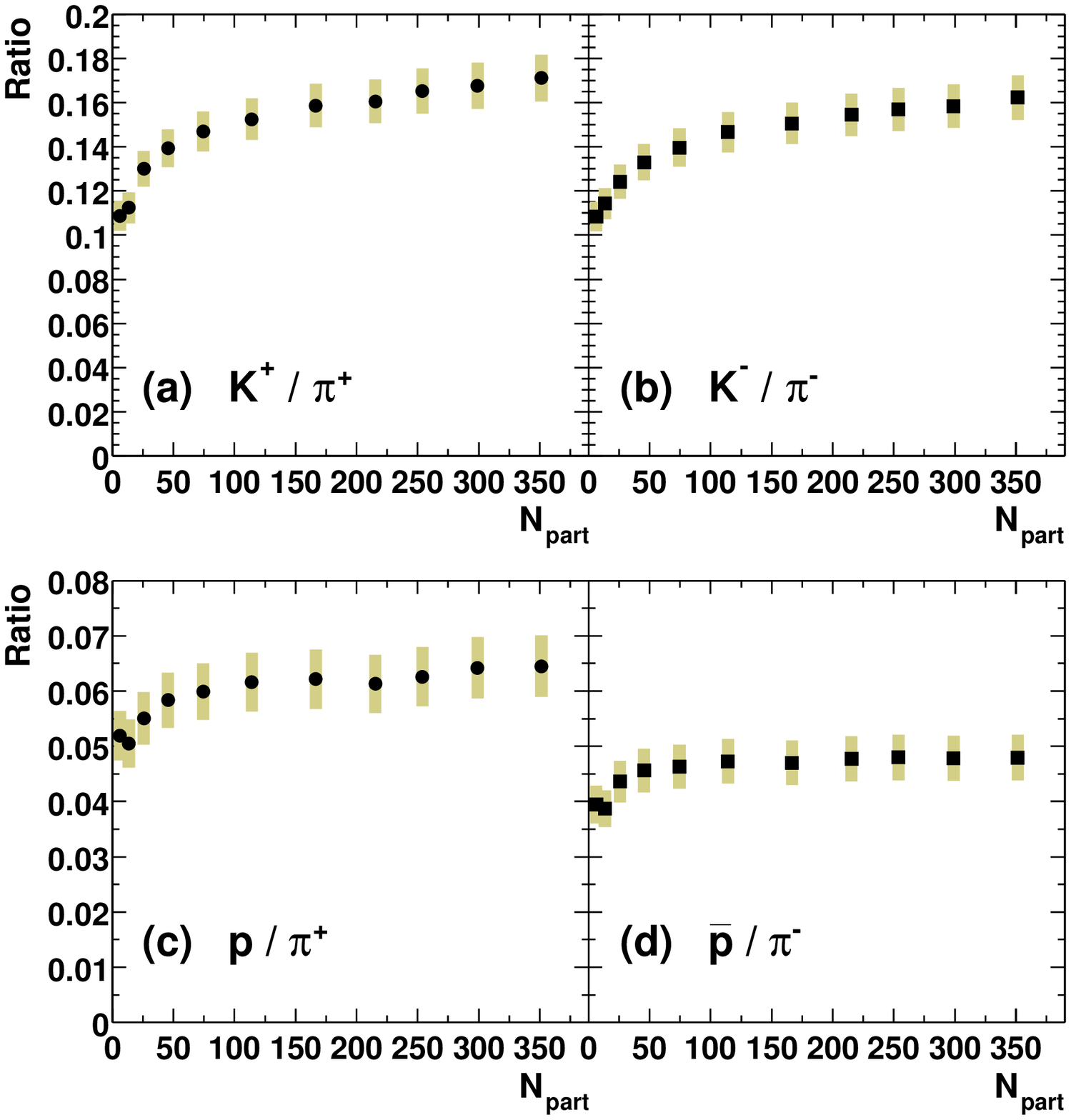, width=0.46\linewidth}&\hspace*{-0.05\linewidth}
\psfig{file=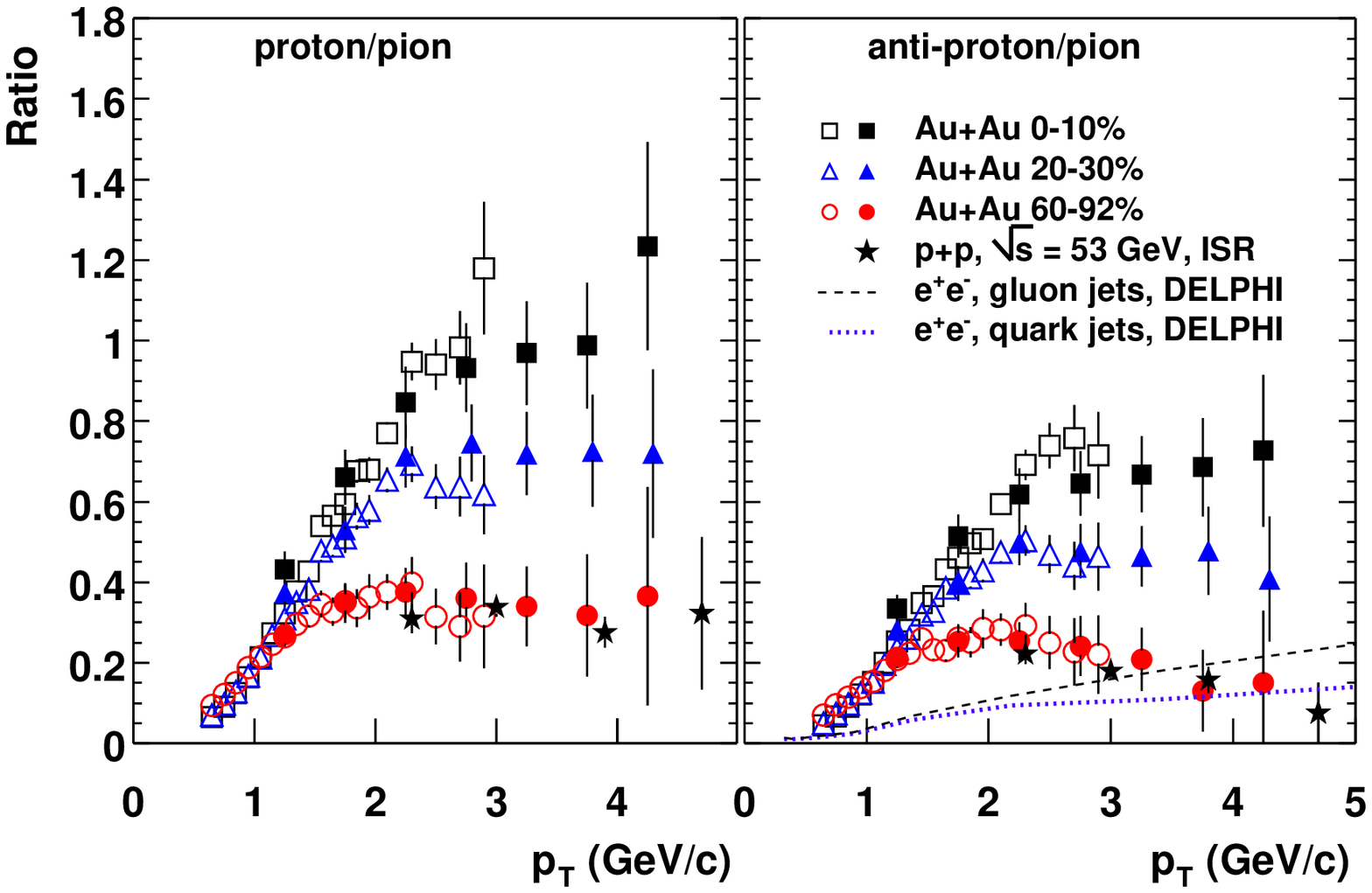,width=0.50\linewidth,height=0.42\linewidth}

\end{tabular}
\end{center}\vspace*{-0.25in}
\caption[]{a) (left) $K^{\pm}/\pi^{\pm}$ and $p^{\pm}/\pi^{\pm}$ ratios as a function of centrality ($N_{\rm part}$) in $\sqrt{s_{NN}}$ Au+Au collisions~\cite{PXWP}. b) (right) $p/\pi$ and $\bar{p}/\pi$ ratio as a function of $p_T$ from the same data~\cite{PXWP,PXscalingPRL91}. Open (filled) points are for $\pi^{\pm}$ ($\pi^0$), respectively. \label{fig:ratios}}

\end{figure}

\begin{figure}[!thb]
\begin{center}
\begin{tabular}{cc}

\psfig{file=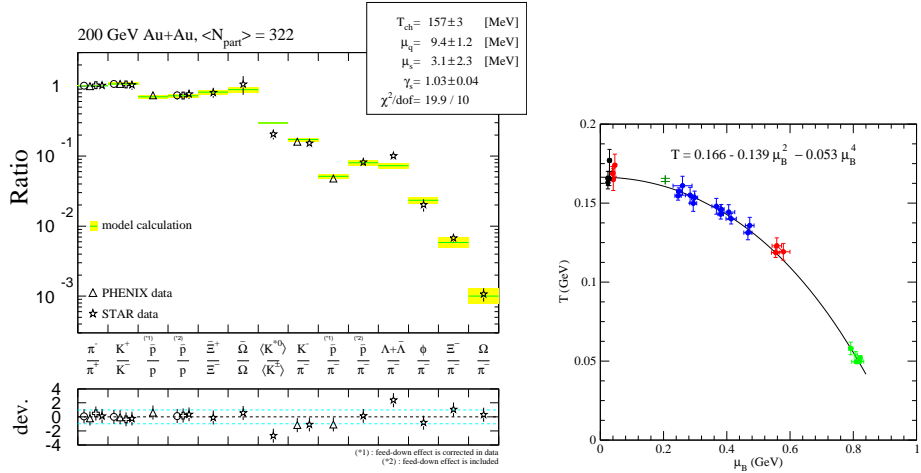,width=0.56\linewidth}&
\psfig{file=figs/Cleymans-TmuB06.eps,width=0.38\linewidth}
\end{tabular}
\end{center}\vspace*{-0.25in}
\caption[]{a)(left) Ratios of $p_T$-integrated mid-rapidity yields for different hadron species measured in Au+Au central central collisions at $\sqrt{s_{NN}}=200$ GeV~\cite{KanetaXu}. BRAHMS(circles) and PHOBOS (crosses) data are shown as well as the PHENIX (triangles) and STAR (stars) data indicated. b) (right) $T_{ch}$ versus $\mu_B$ for thermal model fits as a function of $\sqrt{s_{NN}}$~\cite{CleymansORWheaton}, where the line represents a parameterization of the freeze-out curve, the temperature where chemical equilibrium is achieved so that the particle abundances are fixed. \label{fig:thermalmodels}}

\end{figure}

     The semi-inclusive ratios of different particle abundances also vary smoothly as a function of centrality in Au+Au collisions at RHIC (see Fig.~\ref{fig:ratios}a) with a considerably larger increase in $K^{\pm}$ production than $p$, $\bar{p}$ production relative to $\pi^{\pm}$. 
  However the $p/\pi^{+}$ and $\bar{p}/\pi^{-}$ ratios as a function of $p_T$ (Fig.~\ref{fig:ratios}b) show a dramatic increase as a function of centrality at RHIC~\cite{PXscalingPRL91} which was totally unexpected and is still not fully understood (see below). 
  The ratios of particle abundances (which are dominated by low $p_T$ particles) for central Au+Au collisions at RHIC, even for strange and multi-strange particles, 
are well described (Fig.~\ref{fig:thermalmodels}a) by fits to a thermal distribution, 
          \begin{equation}
{{d^2\sigma} \over {dp_L p_T dp_T}}\sim e^{-(E-\mu)/T} \rightarrow {\bar{p} \over p}=\frac{e^{-(E+\mu_B)/T}}{e^{-(E-\mu_B)/T}}=e^{-(2\mu_B)/T} \qquad ,
\label{eq:boltz2}
\end{equation}
 with similar expressions for strange particles, where $\mu_B$ and $\mu_S$ are chemical potentials associated with each conserved quantity: baryon number ($\mu_B$) and strangeness ($\mu_S$). 
 This should not be very surprising as the particle abundances in A+A collisions at SPS and AGS energies~\cite{PBMStachelNPA606} and in p-p~\cite{BecattiniHeinz} and $e^+ e^-$ collisions~\cite{BecattiniZPC76} are also well described by the same thermal model to such an extent that the chemical freezeout temperature $T_{ch}$ as a function of $\mu_B$ (from Eq.~\ref{eq:boltz2}) could be derived, which looks suspiciously like a phase diagram (Fig.~\ref{fig:thermalmodels}b). Of course, as the thermal equilibrium properties of the QGP are the subject of interest, it is important to understand how or if the thermal properties of the observed hadrons relate to the thermal properties of quarks and gluons in the QGP before getting too excited. Also, while these thermal models appear to be simple they have many technical details which are beyond the scope of this article.

\section{Flow}
   A distinguishing feature of A+A collisions compared to either p-p or p+A collisions is the collective flow observed. This effect is seen over the full range of energies studied in heavy ion collisions, from incident kinetic energy of $100A$ MeV to c.m. energy of $\sqrt{s_{NN}}=200$ GeV~\cite{LaceyQM05}. Collective flow, or simply flow, is a collective effect which can not be obtained from a superposition of independent N-N collisions.

   \begin{figure}[!thb]
\begin{center}

\psfig{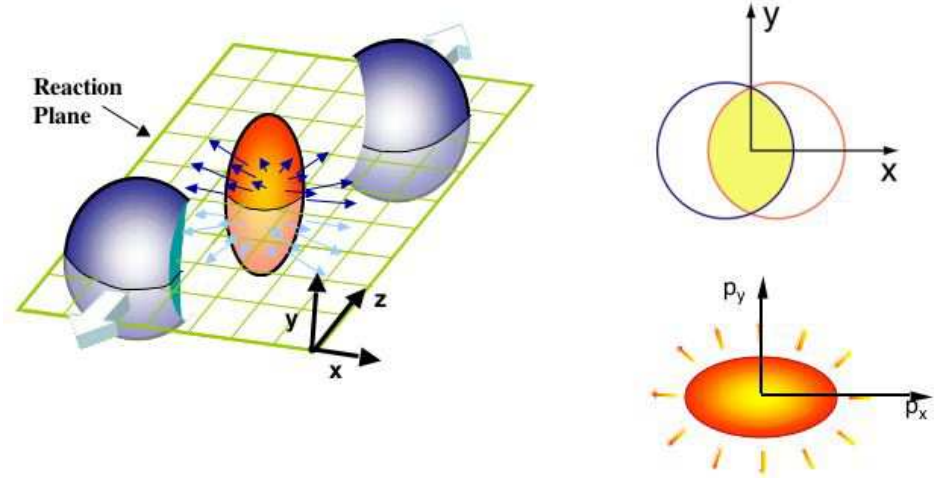}
\end{center}\vspace*{-0.25in}
\caption[]{a) (left) Almond shaped overlap zone generated just after an A+A collision where the incident nuclei are moving along the $\pm z$ axis, and the reaction plane, which by definition contains the impact parameter vector (along the $x$ axis) Thanks to Masashi Kaneta for the figure~\cite{KanetaQM04}. b) (right) View of the collision down the $z$ axis: (top) spatial distribution (bottom) momentum distribution after elliptic flow ($v_2$) develops 
\label{fig:MasashiFlow}}

\end{figure}
   Immediately after an A+A collision, the overlap region defined by the nuclear geometry is almond shaped (see Fig~\ref{fig:MasashiFlow}) with the shortest axis along the impact parameter vector. Due to the reaction plane breaking the $\phi$ symmetry of the problem, the semi-inclusive single particle spectrum is modified from Eq.~\ref{eq:siginv} by an expansion in harmonics~\cite{Ollitrault} of the azimuthal angle of the particle with respect to the reaction plane, $\phi-\Phi_R$~\cite{HeiselbergLevy}, where the angle of the reaction plane $\Phi_R$ is defined to be along the impact parameter vector, the $x$ axis in Fig.~\ref{fig:MasashiFlow}: 
  \begin{eqnarray}
{Ed^3 N \over dp^3}&=&{d^3 N\over p_T dp_T dy d\phi}\\[0.2cm]
&=&{d^3 N\over 2\pi\, p_T dp_T dy} [1+2 v_1 \cos(\phi-\Phi_R)+ 2 v_2\cos2(\phi-\Phi_R)+ \cdots ] .\nonumber
\label{eq:siginv2} 
\end{eqnarray} 
The expansion parameter $v_1$ is called the directed flow and $v_2$ the elliptical flow. If no collective behavior takes place, i.e. the interaction is merely a superposition of independent nucleon-nucleon collisions, then the outgoing momentum distribution of the particles would be isotropic in azimuth. However, since the leading participating nucleons in the forward region $+z$ (Fig.~\ref{fig:MasashiFlow}a) will interact with many other nucleons in the ``almond", they will be pushed away from the rest of the participants, into the $+x$ direction, while the $-z$ going participants are pushed towards $-x$. This is what causes the directed flow, $v_1$, which was discovered at the Bevalac~\cite{PlasticBall} and is clearly sensitive to the Equation of State. For instance if one imagines the almond to be composed of billiard balls requiring lots of pressure for a small deformation (hard EOS) a larger $v_1$ would result than if the almond suddenly melts, perhaps turning into a `perfect fluid', with a much softer EOS~\cite{egseeDLLsci}. 

    The same principles apply to $v_2$, the parameter of $\cos 2(\phi-\Phi_R)$, which (unlike $v_1$) doesn't change sign with rapidity, and hence is non-zero at midrapidity. If thermal equilibrium is reached, then the pressure gradient is directed mainly along the direction of the impact parameter ($x$ axis in Fig.~\ref{fig:MasashiFlow}b) and collective flow develops along this direction. If all the particles are approximately at rest in the fluid and thus move with the fluid velocity, the transverse momentum distribution will reflect the fluid profile. Hence the anisotropic spatial distribution is carried over to an anisotropic momentum distribution through the pressure gradient~\cite{Ollitrault}.   
   
    It is important to emphasize that the spatial anisotropy turns into an momentum anisotropy only if the outgoing particles (or partons) interact with each other~\cite{HeiselbergLevy}. Thus the momentum anisotropy is proportional to the spatial anisotropy of the almond, represented by the eccentricity, $\varepsilon=(R^2_y-R^2_x)/(R^2_y +R^2_x)\simeq (R_y-R_x)/(R_y +R_x)$, at the time ($t_0$) of thermalization. This is due to the fact that the mean number of scatterings in the transverse plane is different along the $x$ and $y$ axes~\cite{VoloshinPoskanzer,SorgePRL82,HeiselbergLevy}.  The mean number of scatterings  is proportional to the particle density, $\rho=(1/\pi R_x R_y)\, dn/dy$ (similar to Eq.~\ref{eq:eBj}) times the interaction cross section ($\sigma$) times the distance traversed:
    \begin{equation}
    v_2\propto R_y\,\sigma {1 \over {\pi R_x R_y}}\, {dn \over dy} - 
     R_x\,\sigma {1\over {\pi R_x R_y}}\, {dn \over dy} \propto 
    \varepsilon\,\sigma {1 \over {\pi R_x R_y}}\, {dn \over dy} \qquad ,
    \label{eq:VP}
    \end{equation}
    where $R_x=\sqrt{\mean{x^2}}$, $R_y=\sqrt{\mean{y^2}}$.

\begin{figure}[!htb]
\begin{center}
\begin{tabular}{cc}
\hspace*{-0.02\linewidth}\psfig{file=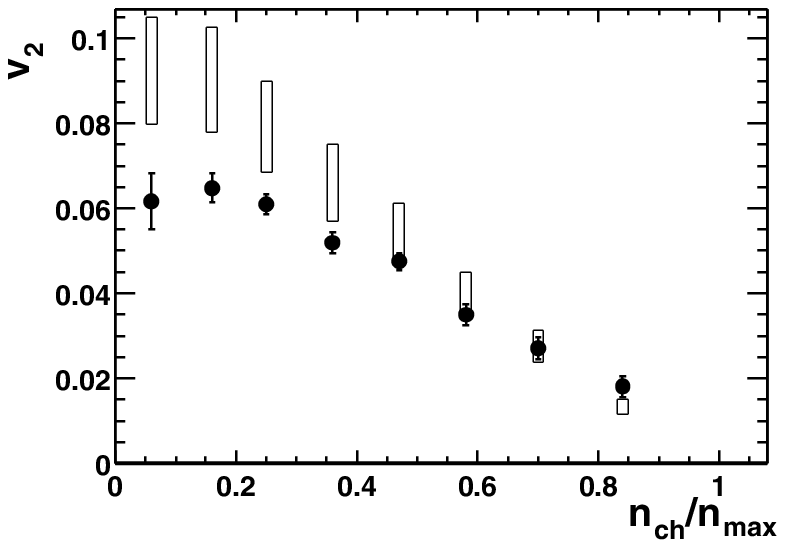,width=0.47\linewidth}&\hspace*{-0.04\linewidth}
\psfig{file=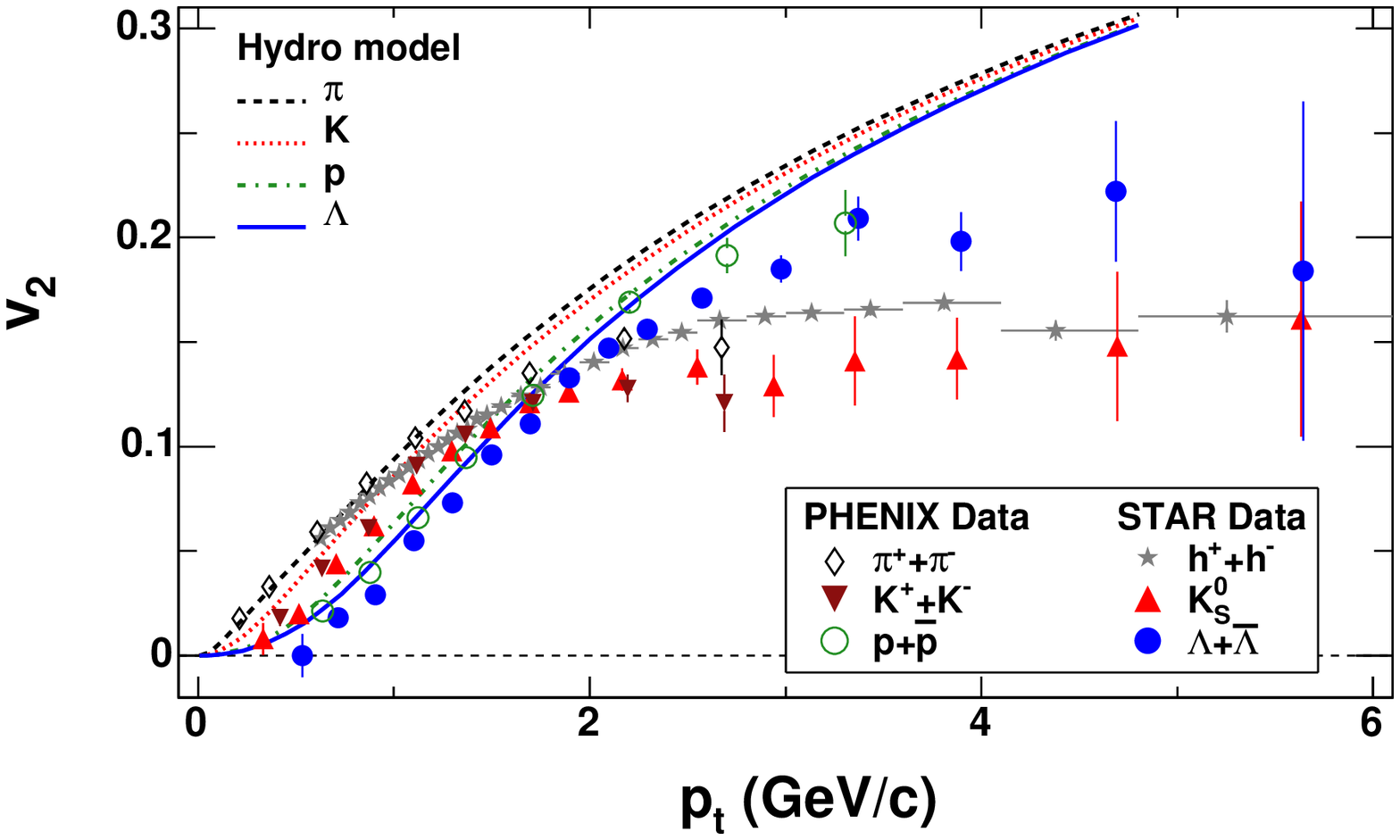,width=0.55\linewidth}
\end{tabular}
\end{center}\vspace*{-0.25in}
\caption[]{a) (left) $v_2$ as function of centrality (measured as a fraction of the total number of charged particles $n_{\rm ch}/n_{max}$)~\cite{STAR130}. The boxes represent the expected hydro-limit with $v_2/\varepsilon=0.19$ (lower edge) and 0.25 (upper edge). b) (right) $v_2$ as a function of $p_T$ for identified particles in minimum bias Au+Au collisions at $\sqrt{s_{NN}}=200$ GeV together with a hydro calculation~\cite{STARPRC72}. \label{fig:flow2}}
\end{figure}

    Since the eccentricity $\varepsilon$ is much larger for peripheral than for central collisions, the dependence of $v_2$ on centrality should exhibit a characteristic shape (see Fig.~\ref{fig:flow2}a)~\cite{STAR130}. This was one of the first publications from RHIC and showed that $v_2$ was surprisingly large and near the hydro-limits.  The hydro-limits indicated are for full thermalization of the system at the value of $\varepsilon$ given by the initial nuclear geometry of the almond at the time of overlap. If the system doesn't thermalize rapidly, the flow tends to vanish because the eccentricity reduces as the system expands~\cite{PXWP,KolbPLB459}. 
    
\begin{figure}[!htb]
\begin{center}
\begin{tabular}{cc}
\psfig{file=figs/TeaneyPRC68.epsf,width=0.39\linewidth}&
\psfig{file=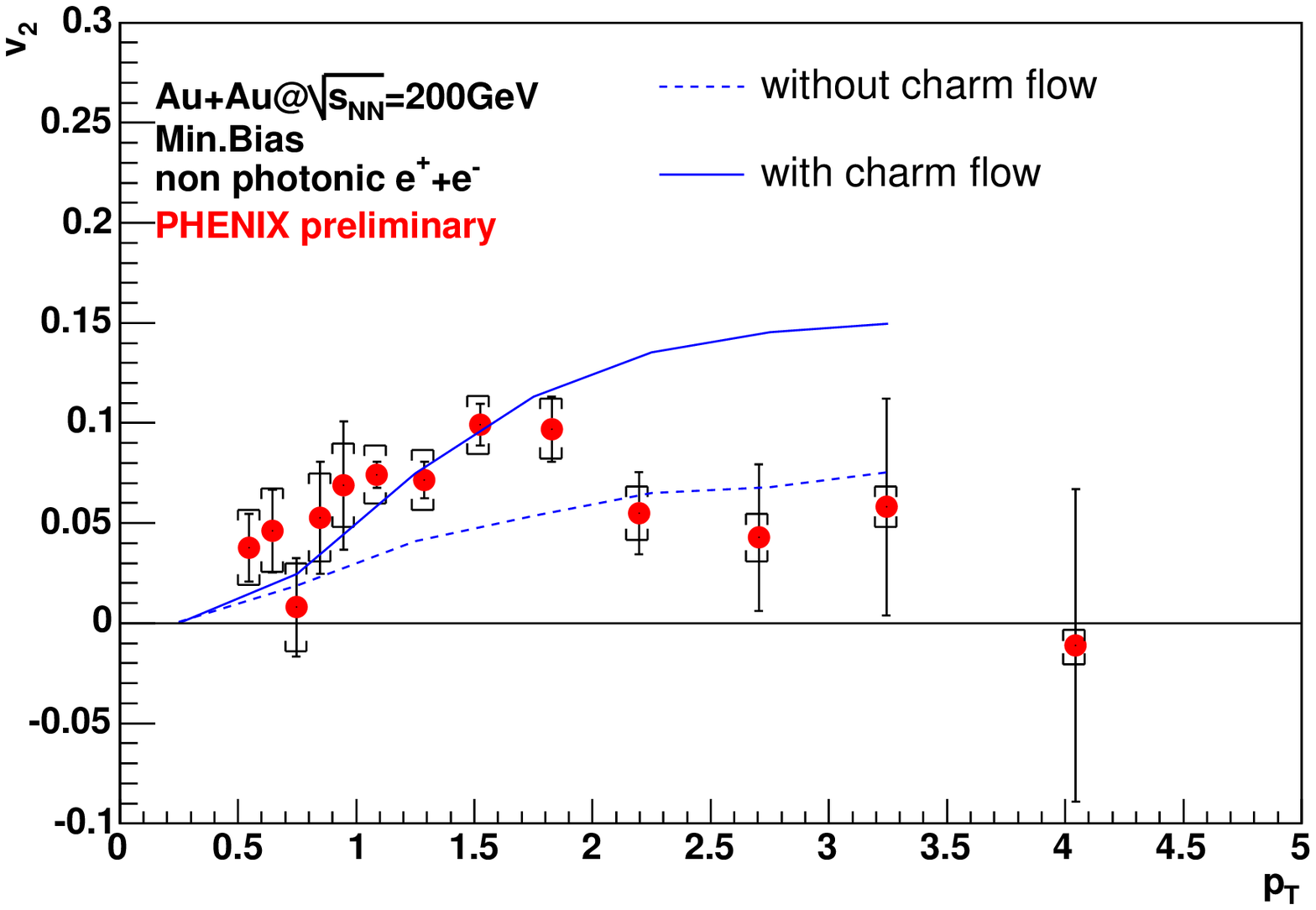,width=0.59\linewidth}
\end{tabular}
\end{center}\vspace*{-0.25in}
\caption[]{a) (left) $v_2$ as a function of $p_T$ in a hydro calculation~\cite{TeaneyPRC68} for mid-central collisions for different values of $\Gamma_s/\tau_0$, the ``sound attenuation length" which is zero for a ``perfect fluid" and increases linearly with the viscosity. b) (right) $v_2$ of non-photonic $e^+ + e^-$ from semi-leptonic heavy flavor decay~\cite{PXcharmAA06}. The solid curve is if both charm and light quarks flow, while the dashed curve is if only the light quark flows~\cite{GrecoKoRapp}. \label{fig:whatflows}}
\end{figure}

    Another surprise~\cite{STARPRL90} (Fig.~\ref{fig:flow2}b)~\cite{STARPRC72} was that the $v_2$ followed the hydro prediction out to $p_T\sim 2$ GeV/c and then plateaued at a constant value to much higher $p_T$. This was one of the principal arguments for the ``perfect fluid" because any modest value of viscosity~\cite{TeaneyPRC68} would cause the $v_2$ to decrease towards zero near $p_T\sim 1.7$ GeV/c (see Fig.~\ref{fig:whatflows}a, below). 

\begin{figure}[!htb]
\begin{center}
\begin{tabular}{cc}
\psfig{file=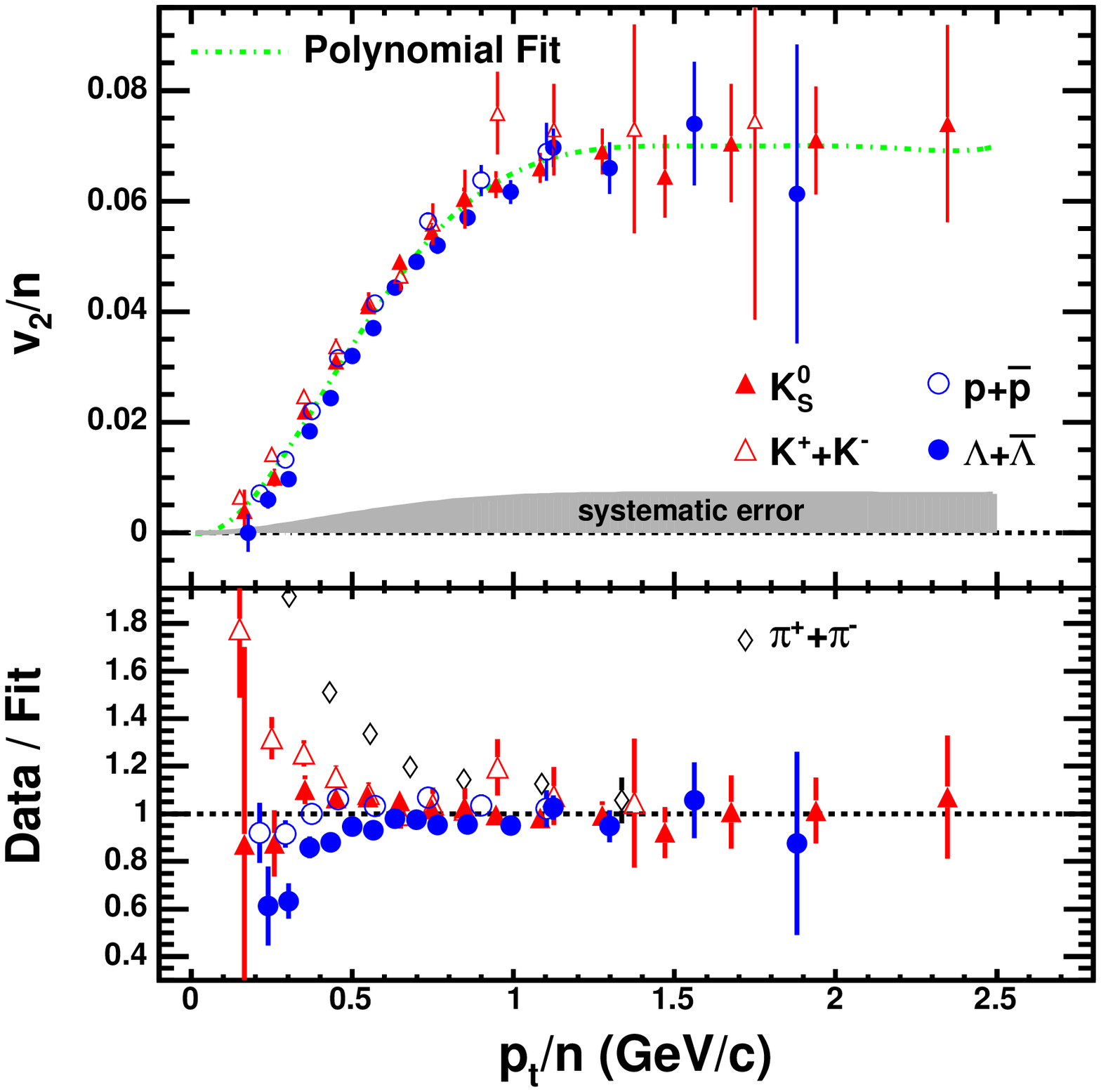,width=0.5\linewidth}&\hspace*{-0.044\linewidth}
\psfig{file=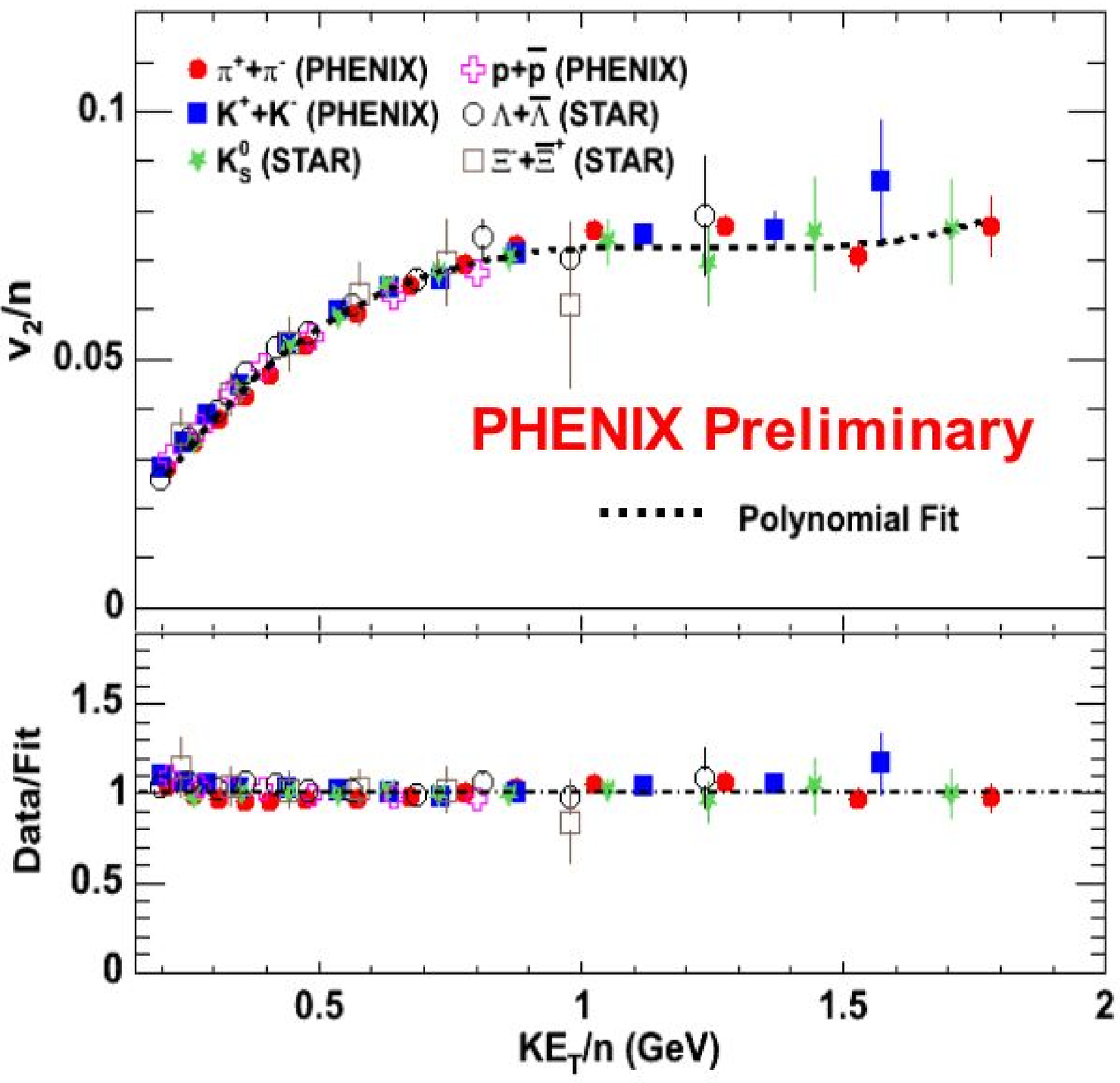,width=0.49\linewidth}
\end{tabular}
\end{center}\vspace*{-0.15in}
\caption[]{a) (left) $v_2/n$ vs $p_T/n$ for identified particles, where $n$ is the number of constituent quarks, together with a polynomial fit~\cite{STARPRC72}. Bottom panel shows deviations from the fit. b) (right) $v_2/n$ vs transverse kinetic energy per constituent quark $KE_{T}/n$~\cite{PXArkadyQM06}.    \label{fig:flow3}}
\end{figure}
	As hydrodynamics appears to work in both p-p and A+A collisions, and collective flow is observed in A+A collision over the full range of energies studied, a key question is what is flowing at RHIC and is it qualitatively different from the flow observed at lower $\sqrt{s_{NN}}$ ? One interesting proposal in this regard is that the constituent quarks flow~\cite{VoloshinQM02}, so that the flow should be proportional to the number of constituent quarks $n_q$, in which case $v_2/n_q$ as a function of $p_T/n_q$ would represent the constituent quark flow as a function of constituent quark transverse momentum and would be universal.  
Interestingly, the RHIC data~\cite{STARPRC72} (Fig.~\ref{fig:flow3}a) seem to support this picture, although the fact that the $\pi^+ +\pi^-$ deviate most from the universal curve should raise some suspicions as the pion is the only particle whose mass is much less than that of its constituent quarks. However, in relativistic hydrodynamics, at mid-rapidity, the transverse kinetic energy, $m_T-m_0=(\gamma_T-1) m_0\equiv KE_T$, rather than $p_T$ is the relevant variable, and in fact $v_2/n_q$ as a function of $KE_T/n_q$ seems to exhibit nearly perfect scaling~\cite{PXArkadyQM06} (Fig.~\ref{fig:flow3}b).

Another striking hint as to what is flowing at RHIC is given in Fig.~\ref{fig:whatflows}b where charm particles detected by their large semi-leptonic decay~\cite{CCRS,PXPRL88} exhibit the same $v_2$ as other particles~\cite{PXPRC72charm,PXcharmAA06}. For the charm particles, the $v_2$ of the decay electrons follows the $v_2$ of the $D (c\bar{d},c\bar{u})$ mesons~\cite{Batsouli,GrecoKoRapp}, but due to their different masses the $c$ and $\bar{u}, \bar{d}$ quarks have different momenta for the same velocity required for formation by coalescence, so the $v_2$ of the $D$ mesons is reduced to that of the light quark at lower $p_T$ if the $c$ quark itself does not flow. The data (Fig.~\ref{fig:whatflows}b) favor the flow of the $c$ quark, but clearly lots more work remains to be done to improve both the measurement and the theory. However since the $\phi (s\bar{s})$ meson also flows at RHIC~\cite{PXphiQM05} and follows the constituent quark scaling rule, this clearly indicates that the flow at RHIC is partonic because the hadronic interaction cross section of the $\phi$ meson is much smaller than for other hadrons~\cite{Lipkin,Joos,KT67}. 

\section{Jet Quenching}

   A new tool for probing the color response function of the medium was developed in the early 1990's~\cite{Gy1} and given a firm basis in QCD~\cite{BSZARNS00} just before RHIC turned on. In the initial collision of the two nuclei in an A+A collision, when the Lorentz contracted nuclei are overlapped, hard scatterings can occur which produce pairs of outgoing high $p_T$ partons with their color charge fully exposed as they travel through the  medium before they fragment to jets of particles. If the medium also has a large density of color charges, then the partons are predicted to lose energy by `coherent' (LPM) gluon bremsstrahlung which  is sensitive to the properties of the medium. This leads to a reduction in the $p_T$ of the partons and their fragments and hence a reduction in the number of partons or fragments at a given $p_T$, which is called jet quenching. The effect should be  absent in p+A or d+A collisions due to the lack of a medium produced. 
   
\begin{figure}[!htb]
\begin{center}
\begin{tabular}{cc}
\psfig{file=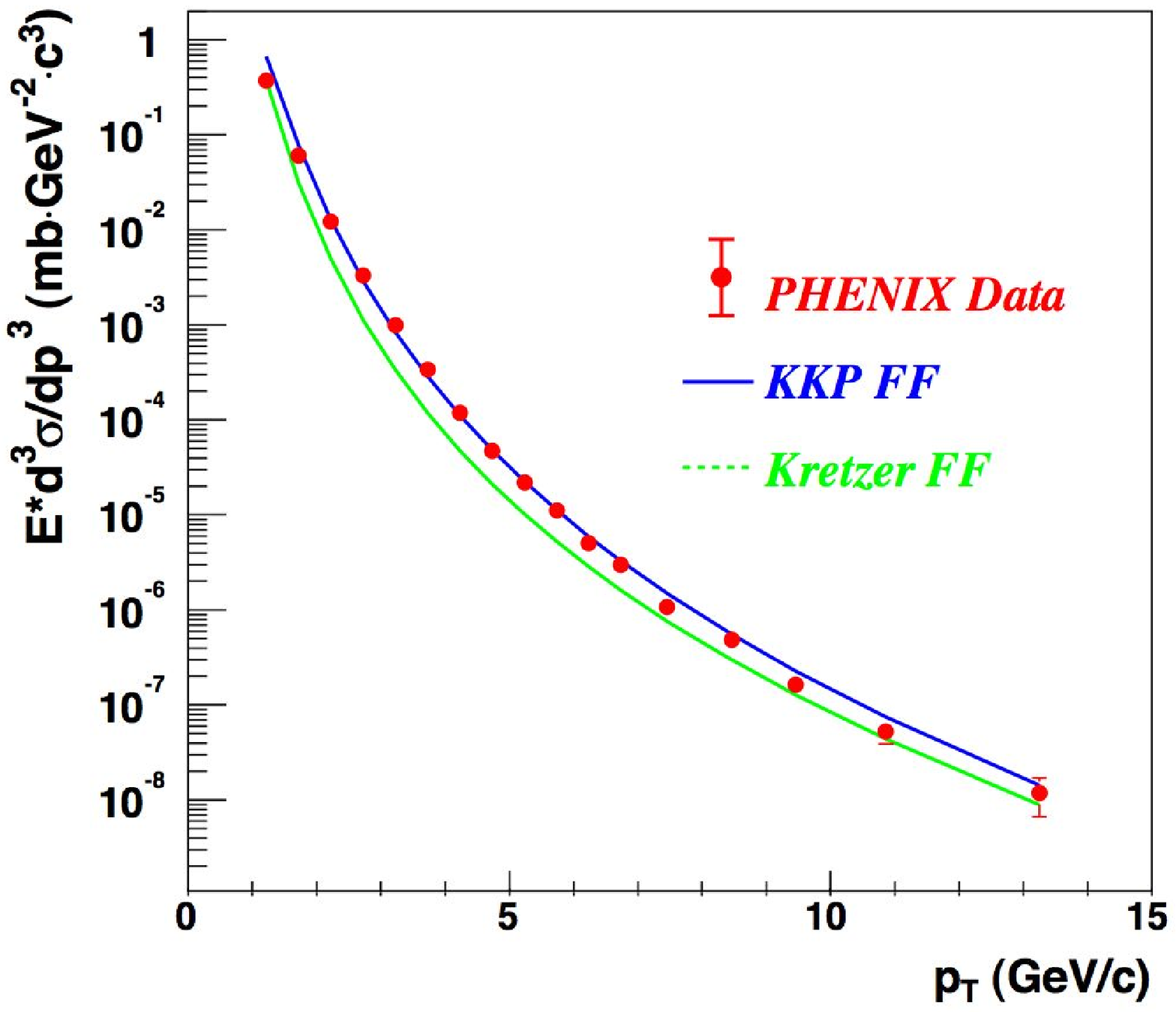,width=0.48\linewidth,height=0.44\linewidth}&
\hspace*{-0.02\linewidth}\psfig{file=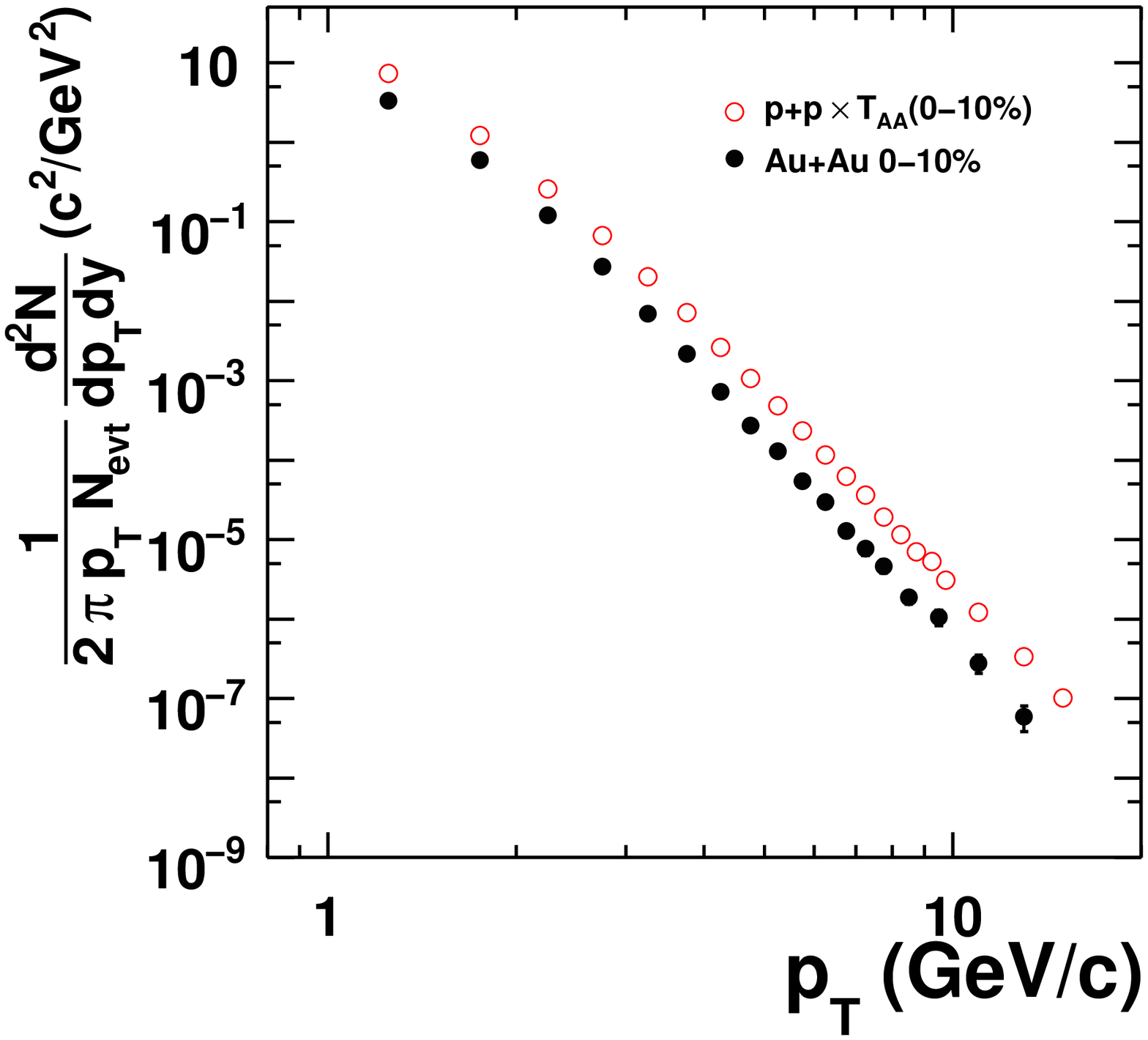,width=0.48\linewidth}
\end{tabular}
\end{center}\vspace*{-0.25in}
\caption[]{a)(left) Invariant cross section~\cite{PXpppi0} of $\pi^0$ at mid-rapidity from p-p collisions at $\sqrt{s_{NN}}=200$~GeV together with NLO pQCD predictions from Vogelsang~\cite{WV} with two different fragmentation functions. b) (right) p-p data from a) multiplied by $\mean{T_{AA}}$ for Au+Au central collisions (1-10\%) plotted on a log-log scale together with the measured~\cite{PXppg054} semi-inclusive $\pi^0$ invariant yield in Au+Au central (0-10\%) collisions at $\sqrt{s_{NN}}=200$ GeV.    \label{fig:PXpi0spectra}}
\end{figure}

    One of the major, arguably 
{\it the} major discovery at RHIC, was the observation of jet quenching~\cite{PXppg003,seealsoQM01} by the suppression of $\pi^0$ and non-identified charged hadrons in Au+Au collisions at mid-rapidity for large  transverse momenta, $p_T > 2$ GeV/c. In p-p collisions, particles with $p_T\geq 2$ GeV/c at mid-rapidity (perpendicular to the collision axis) are produced from states with two roughly back-to-back jets which are the result of hard-scattering of the constituents of the nucleon (current-quarks and gluons) as described by QCD. The suppression of high $p_T$ particles  was observed by all four RHIC  experiments~\cite{PXWP,BRWP,PHWP,STWP} and is well calibrated by using measurements in p-p, d+Au and Au+Au collisions at $\sqrt{s_{NN}}=200$ GeV.   
    The $p_T$ spectra of $\pi^0$ from p-p collisions at $\sqrt{s}=200$ GeV~\cite{PXpppi0} and Au+Au central collisions at $\sqrt{s_{NN}}=200$ GeV~\cite{PXpi0AuAu200,PXppg054} are shown in Fig.~\ref{fig:PXpi0spectra}.  
Note that the p-p measurements for $p_T \geq 3$ GeV/c are in excellent agreement with a pQCD calculation~\cite{WV} and that the $p_T$ spectra in both p-p and central Au+Au follow a pure power law for $p_T > 3$ GeV/c.

    Since hard scattering is point-like, with distance scale $1/p_T< 0.1$ fm, the cross section in p+A (B+A) collisions, compared to p-p, should be simply proportional to the relative number of possible point-like encounters~\cite{MMay}, a factor of $A$ ($BA$) for p+A (B+A) minimum bias collisions. For semi-inclusive reactions in centrality class $f$ at impact parameter $b$, the scaling is proportional to $\mean{T_{AB}}_{f}$, the overlap integral of the nuclear thickness functions~\cite{Vogt99}  averaged over the centrality class $f$. Since $\mean{T_{AB}}_f=\langle N_{coll}\rangle_f/\sigma_{NN}$, where $\langle N_{coll}\rangle_f$ is the average number of binary nucleon-nucleon 
inelastic collisions with cross section $\sigma_{NN}$ for the centrality class $f$, point-like scaling is often called binary-collision (or $N_{coll}$)-scaling. However,  the scaling has nothing to do with the inelastic hadronic collision probability, it is proportional only to the geometrical factor $\mean{T_{AB}}_{f}$. 

    Effects of the nuclear medium, either in the initial or final state, may modify the point-like scaling. This is shown rather dramatically in Fig.~\ref{fig:PXpi0spectra}b, where the Au+Au central (0-10\%) collision data are suppressed relative to the scaled p-p data by a factor of $\sim 4-5$ for $p_T\geq 3$ GeV/c. 
    While the suppression of $\pi^0$ at a given $p_T$ in Au+Au compared to the scaled p-p spectrum may be imagined as a loss of these particles due to, for instance, the stopping or absorption of a certain fraction of the parent partons in an opaque medium, it is evident from Fig.~\ref{fig:PXpi0spectra}b that an equally valid representation can be given by a downshift of the scaled p-p spectrum due to, for instance, the energy loss of the parent partons in the medium. 
    
    A quantitative evaluation of the suppression is made using the ``nuclear modification factor'', $R_{AB}$, the ratio of the measured semi-inclusive yield to the point-like scaled p-p cross section: 
\begin{equation}
R_{AB} = \frac{dN_{AB}^P}{\langle T_{AB} \rangle_{f} \times d\sigma_{NN}^P}
       = \frac{dN_{AB}^P}{\langle N_{coll} \rangle_{f} \times dN_{NN}^P} \label{eq:RAB}
\end{equation}
where $dN_{AB}^P$ is the differential yield of a point-like process $P$
in an $A+B$ collision and $d\sigma_{NN}^P$ is the cross section of $P$ in an $NN$ (usually p-p) collision.  For point-like scaling, $R_{AB}=1$. 

\begin{figure}[!thb]
\begin{center}
\begin{tabular}{cc}
\psfig{file=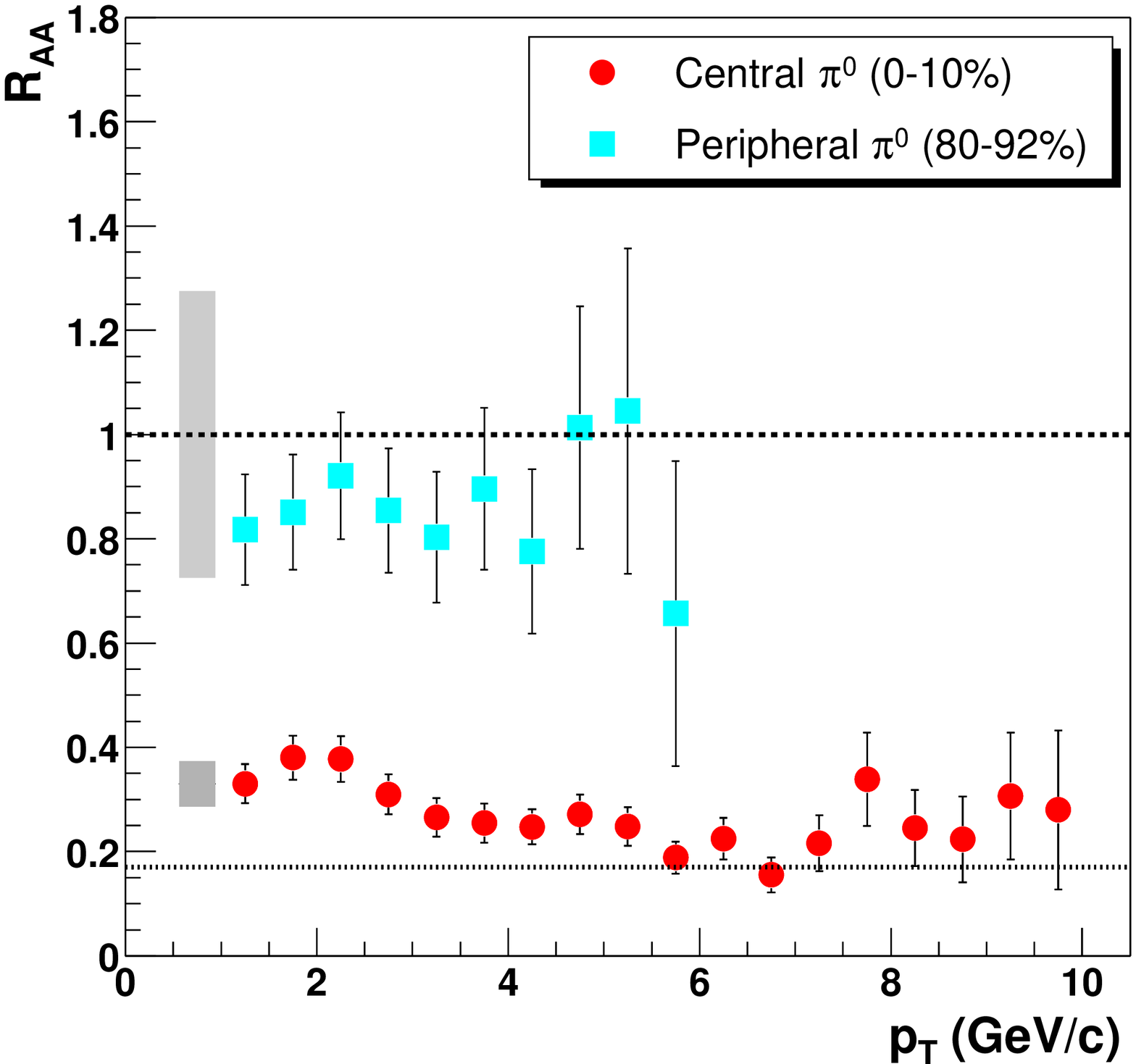,width=0.45\linewidth}&
\psfig{file=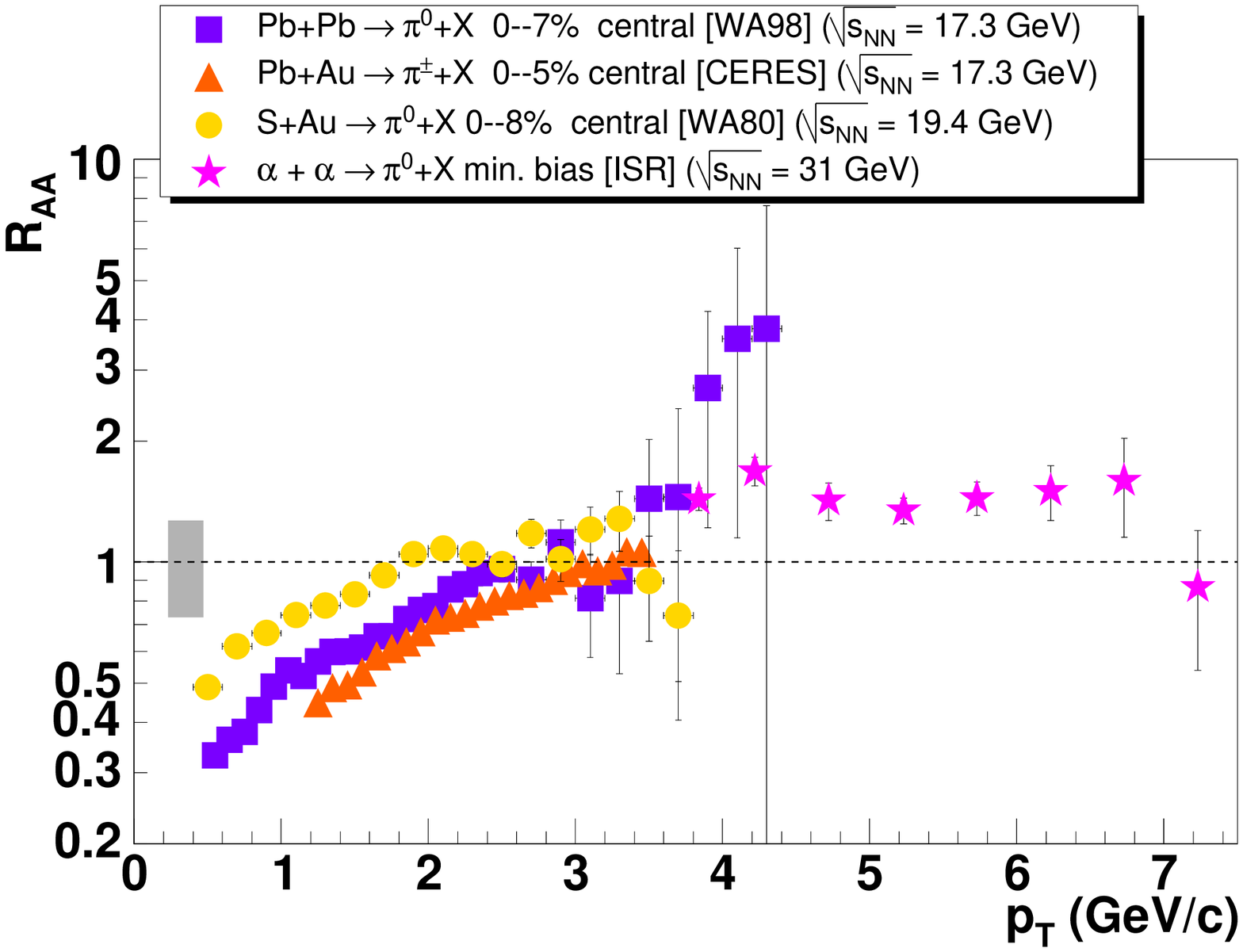,width=0.49\linewidth,height=0.418\linewidth}
\end{tabular}
\end{center}\vspace*{-0.25in}
\caption[]{a) (left) Nuclear modification factor $R_{AA}(p_T)$ for $\pi^0$ in central  and peripheral Au+Au at $\sqrt{s_{NN}}=200$ GeV~\cite{PXpi0AuAu200}. The error bars include all point-to-point experimental (p-p, Au+Au) errors. The shaded bands represent the fractional systematic uncertainties for each centrality which can move all the points at that centrality up and down together. b) (right) Compilation of $R_{AA} (p_T)$ for $\pi^0$ production in A+A collisions from Refs.~\cite{PXWP,DdEPLB04}. \label{fig:RAA1}}
\end{figure}

	Fig.~\ref{fig:RAA1}a shows that there is no suppression of $\pi^0$ in Au+Au peripheral collisions. However, the suppression in Au+Au central collisions, although quite impressive in its own right, is even more dramatic when compared to previous data. All previous measurements of nuclear effects at high $p_T\geq 2$ GeV/c in p+A and A+A collisions at lower $\sqrt{s_{NN}}$ have given results which are larger than point-like scaling (Fig.~\ref{fig:RAA1}b), a situation called the `Cronin Effect'~\cite{Cronin} and thought to be due to the multiple scattering of the incident partons in the nuclear matter before the hard-collision~\cite{Krzy,Lev}. The suppression observed at RHIC is a totally new effect.

\begin{figure}[htb]
\begin{center}
\begin{tabular}{cc}
\psfig{file=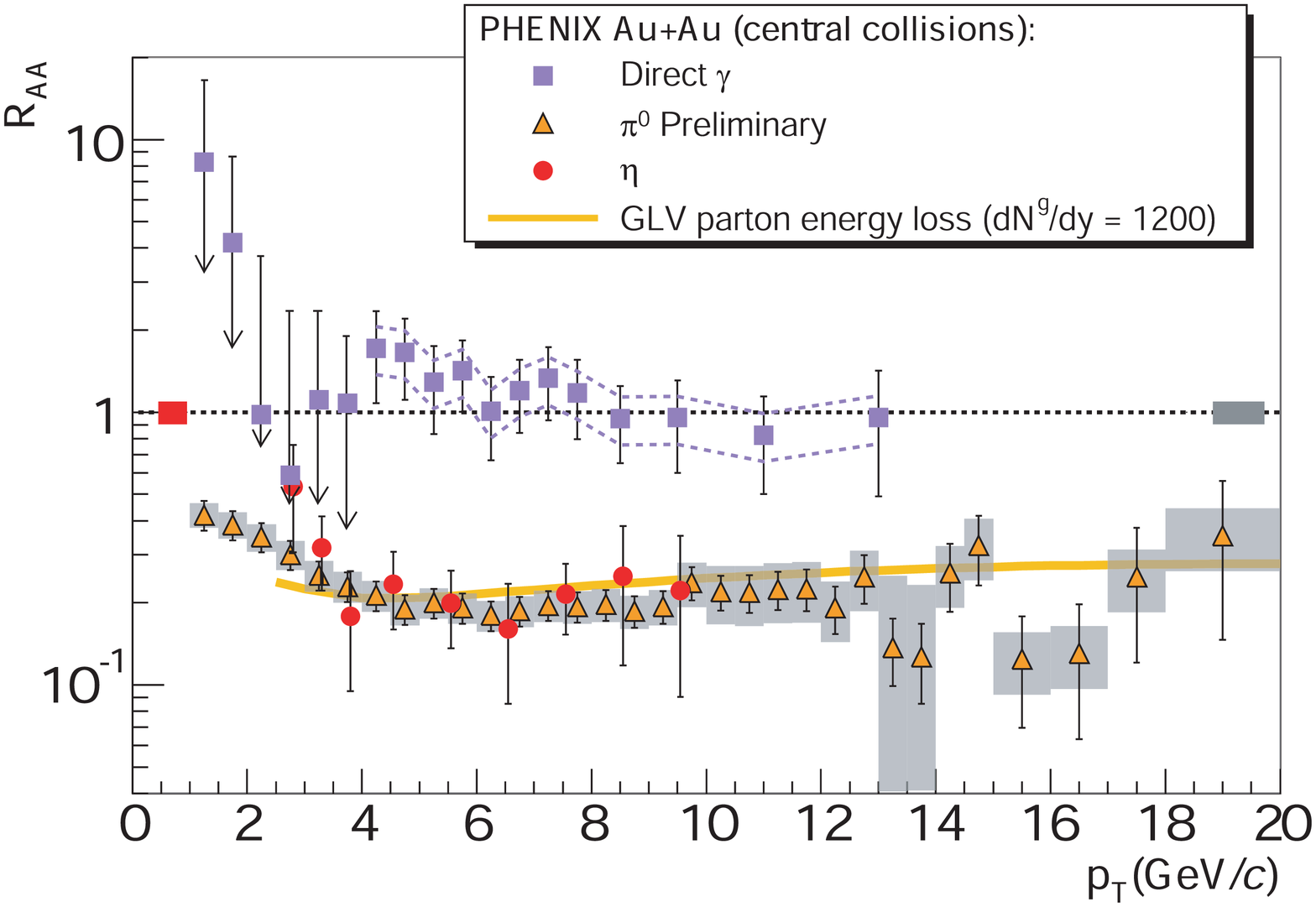,width=0.52\linewidth}&\hspace*{-0.04\linewidth}
\psfig{file=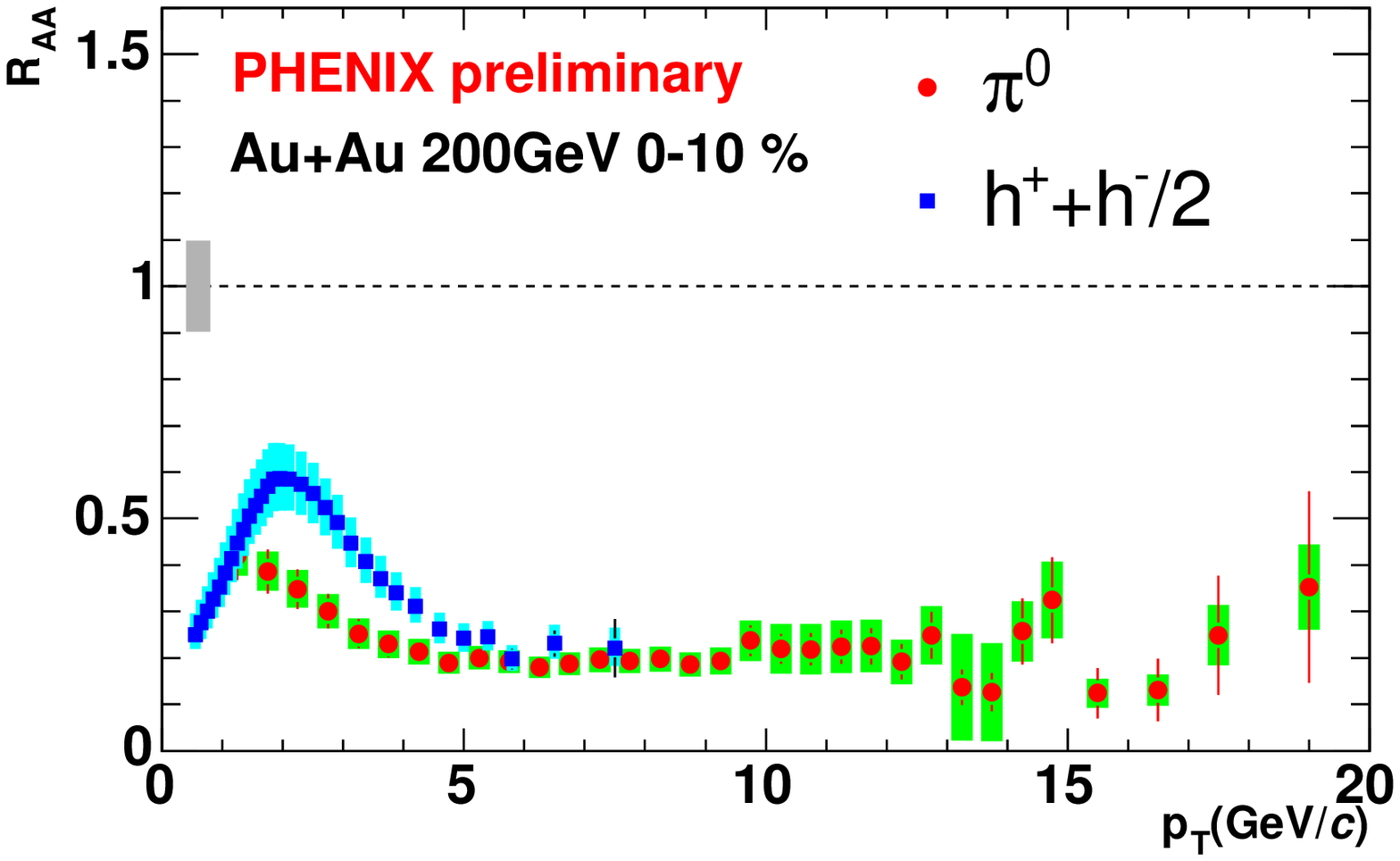,width=0.48\linewidth}
\end{tabular}
\end{center} \vspace*{-0.1in}
\caption[]{a) (left) Nuclear modification factor, $R_{AA}$ for direct photons, $\pi^0$ and $\eta$ in Au+Au central collisions at $\sqrt{s_{NN}}=200$ GeV~\cite{AkibaQM05}. b) $R_{AA}$ for identified $\pi^0$ and-non identified charged hadrons $(h^+ + h^-)/2$ for central (0-10\%) Au+Au collisions at $\sqrt{s_{NN}}=200$ GeV~\cite{MayaQM05}. \label{fig:RAA2}}
\end{figure}

\begin{figure}[!htb]
\begin{center}
\begin{tabular}{cc}
\psfig{file=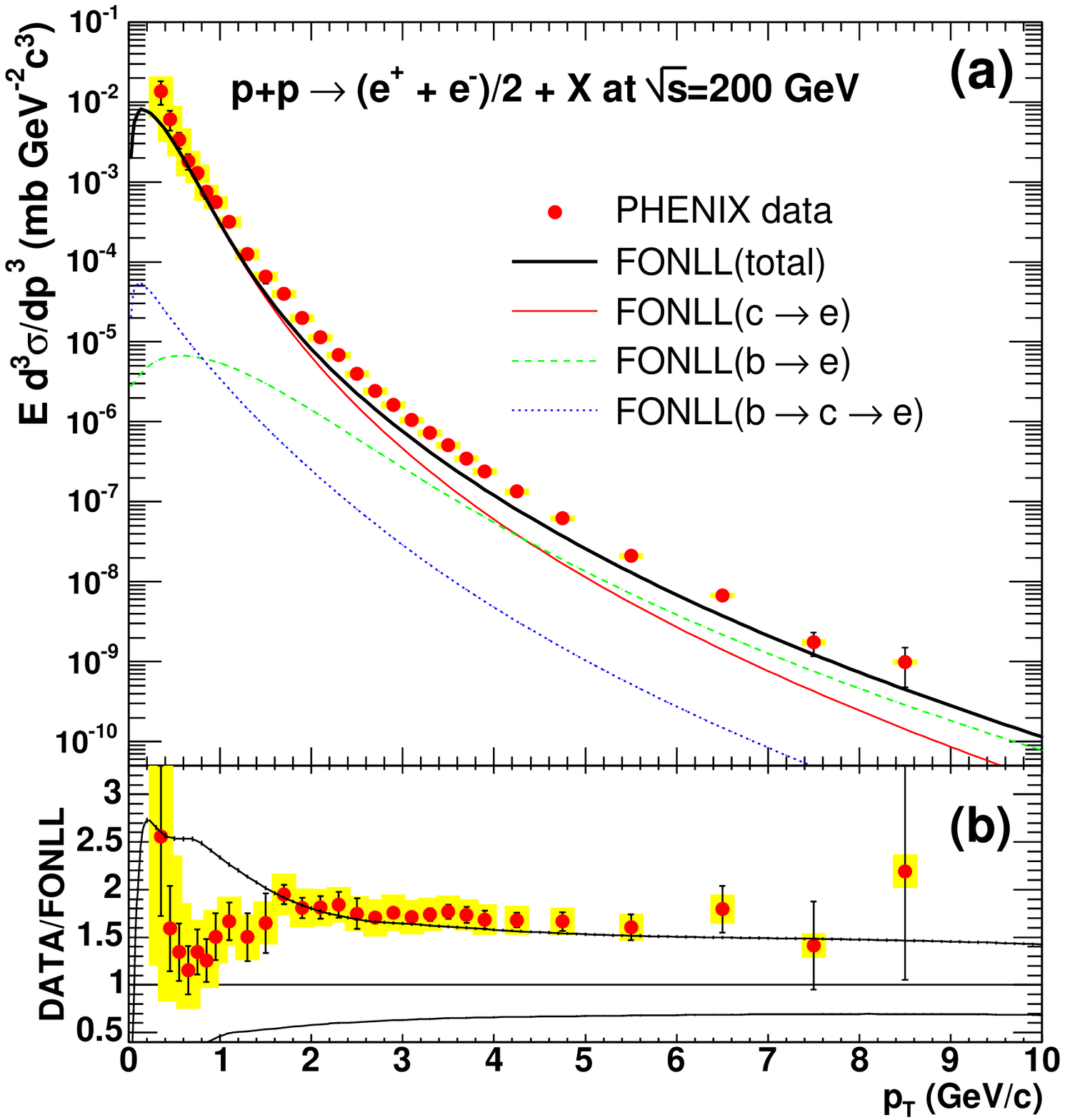,width=0.52\linewidth}&\hspace*{-0.04\linewidth}
\psfig{file=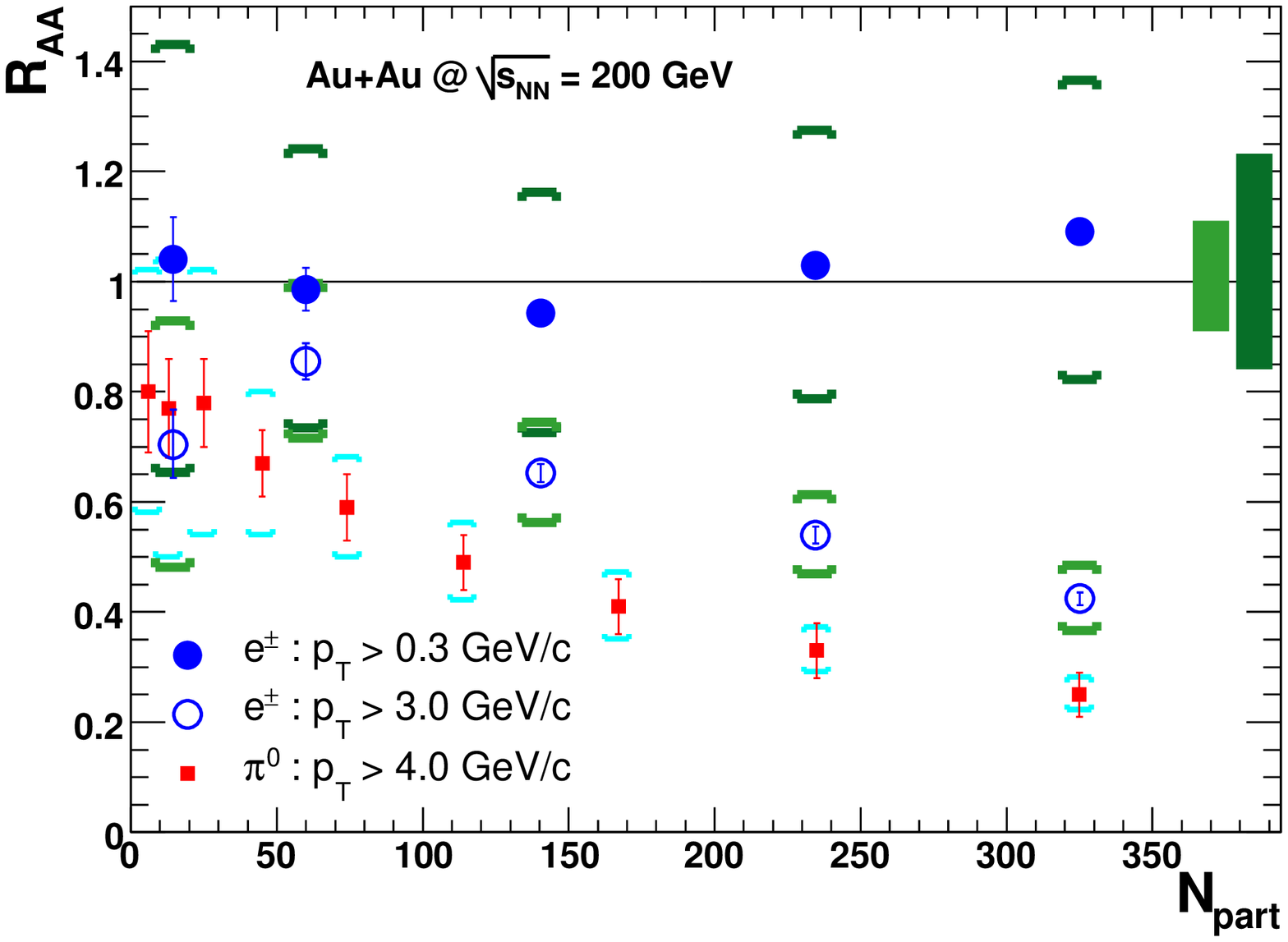,width=0.43\linewidth,height=0.35\linewidth}
\end{tabular}
\end{center}\vspace*{-0.02\linewidth}
\caption[]{a) (left) Invariant cross section~\cite{PXcharmpp06} for prompt (non-photonic) electrons in p-p collisions at $\sqrt{s}=200$ GeV compared to a theoretical calculation (see Ref.~\refcite{PXcharmpp06}) which shows  contributions from both $c$ and $b$ quarks, which depend on $p_T$. b) (right) $R_{AA}$ as a function of centrality ($N_{\rm part}$) for the total yield of $e^{\pm}$ from charm ($p_T > 0.3$) GeV/c, compared to the suppression of the $e^{\pm}$ yield at large $p_T>3.0$ GeV/c which is comparable to that of $\pi^0$ with ($p_T>4$ GeV/c)~\cite{PXcharmAA06}.\label{fig:RAA3}}
\end{figure}

     Naturally, the first question asked about the RHIC suppression was whether it is an initial state effect, produced, for instance,  by `shadowing' of the structure functions in nuclei, or a final state effect produced by the medium. Although originally answered by all 4 RHIC experiments by the observation of no suppression in d+Au collisions~\cite{BRWP,PHWP,STWP,PXWP}, a clearer answer came from later measurements~\cite{PXdirg} of QCD hard-photon production ($gq\rightarrow \gamma q$)~\cite{QCDcompton}, Fig.~\ref{fig:RAA2}a, and the total yield of charm particles ($gg\rightarrow c\bar{c}$)  deduced from measurements~\cite{PXcharmpp06,PXcharmAA06}  of non-photonic $e^{\pm}$ (Fig.~\ref{fig:RAA3}) in p-p and Au+Au collisions.  
     Both these reactions are sensitive to the same initial state partons as $\pi^0$ production, but the photons, which are themselves elementary constituents  which participate in and emerge directly from the hard-scattering, do not interact with the final state medium; and the number of $c \bar{c}$  pairs produced does not depend on whether or not the $c$ or $\bar{c}$ later interact with the medium. Hence the fact that $R_{AA}=1$ as a function of centrality for these two reactions, which is dramatically different from the suppression of $\pi^0$, indicates that the $\pi^0$ suppression is produced by the interaction of the outgoing hard-scattered parton with the medium, losing energy before it fragments into a $\pi^0$. This conclusion is reinforced by the equal suppression of $\eta$-mesons and $\pi^0$ which are both fragments of jets. The curve on Fig.~\ref{fig:RAA2}a shows a theoretical prediction~\cite{VG,MayaQM05} from a model of parton energy loss. The model assumes an inital parton density $dN/dy = 1200$, which corresponds to an energy density of approximately 15 GeV/fm$^3$. The theory curve appears to show a reduction in suppression with increasing $p_T$ while the data appear to be flat to within the errors which clearly could still be improved.

A major surprise was revealed by the measurement of the $p_T$ dependence of non-photonic $e^{\pm}$ from charm particles,  
which exhibits a suppression comparable to $\pi^0$ (Fig.~\ref{fig:RAA3}b), indicating a surprisingly strong interaction with the medium~\cite{PXcharmAA06}. This is is a very recent exciting result which is causing  a detailed reevaluation of the theoretical models because the heavy $c$ quark (and the heavier $b$ quark) were predicted to lose much less energy to the medium than light quarks and gluons~\cite{DjorPRL94}.

Another important result, as indicated in Fig.~\ref{fig:RAA2}b by the difference of $R_{AA}$ of unidentified charged hadrons ($h^+ + h^{-}$) and $\pi^0$  in the range $1.5 < p_T < 5$ GeV/c, is that protons and anti-protons are not suppressed at all in this $p_T$ range, but follow point-like scaling unlike the $\pi^{\pm}$ which are suppressed, which results in an anomalously large $p^{\pm}/\pi^{\pm}$ ratio (Fig.~\ref{fig:ratios}b). The $p^{\pm}/\pi^{\pm}$ ratio returns to the normal value consistent with jet-fragmentation for $p_T\ \gsim\ 5$ GeV/c.  (The $N_{coll}$ scaling of the $p_T$ spectra can be seen in Fig.~\ref{fig:radialflow}b). This observation, called the `baryon anomaly', remains unexplained.

\subsection{Jet properties from two-particle correlations}

    The study of jet properties via two-particle correlations, pioneered at the CERN-ISR~\cite{egseeMJTHP04}, is used at RHIC rather than reconstructing jets from all particles within a cone of size $\Delta r=\sqrt{(\Delta \eta)^2 + (\Delta \phi)^2)}$.  This is because the large multiplicity in A+A collisions (recall Fig.~\ref{fig:collstar}b,c) results in a huge energy $\pi
(\Delta r)^2 \times {1 \over {2\pi}} {{dE_T} \over {d\eta}} \sim 375$ GeV in the standard cone, $\Delta r=1$, for Au+Au central collisions. Elliptical flow further complicates the jet measurement because the large non-jet background is modulated by $\cos2\phi$, which is comparable in width to the jets studied so far. 
\begin{figure}[!thb]
\begin{center}
\begin{tabular}{cc}
\psfig{file=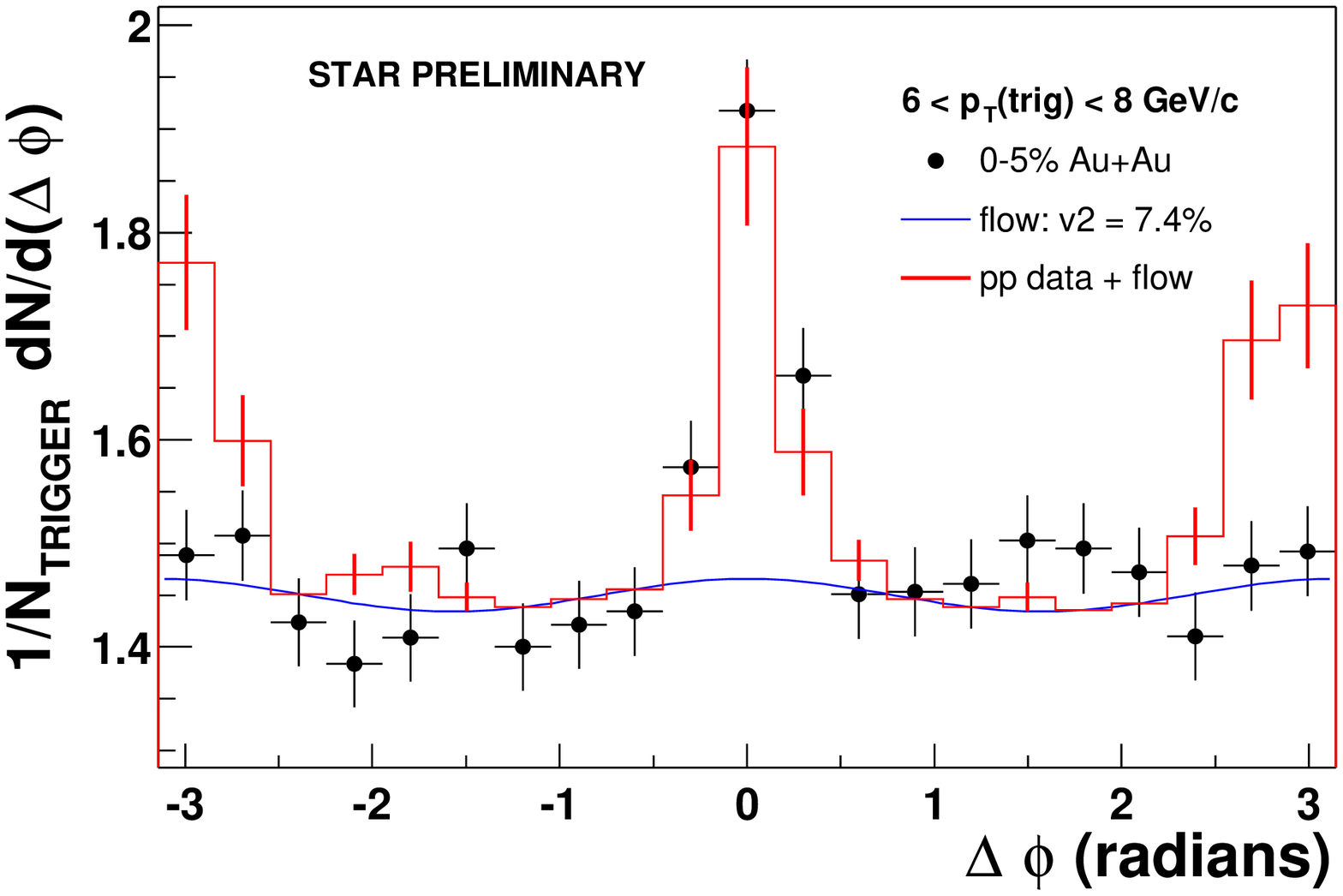,width=0.52\linewidth}& \hspace*{-0.03\linewidth} 
\psfig{file=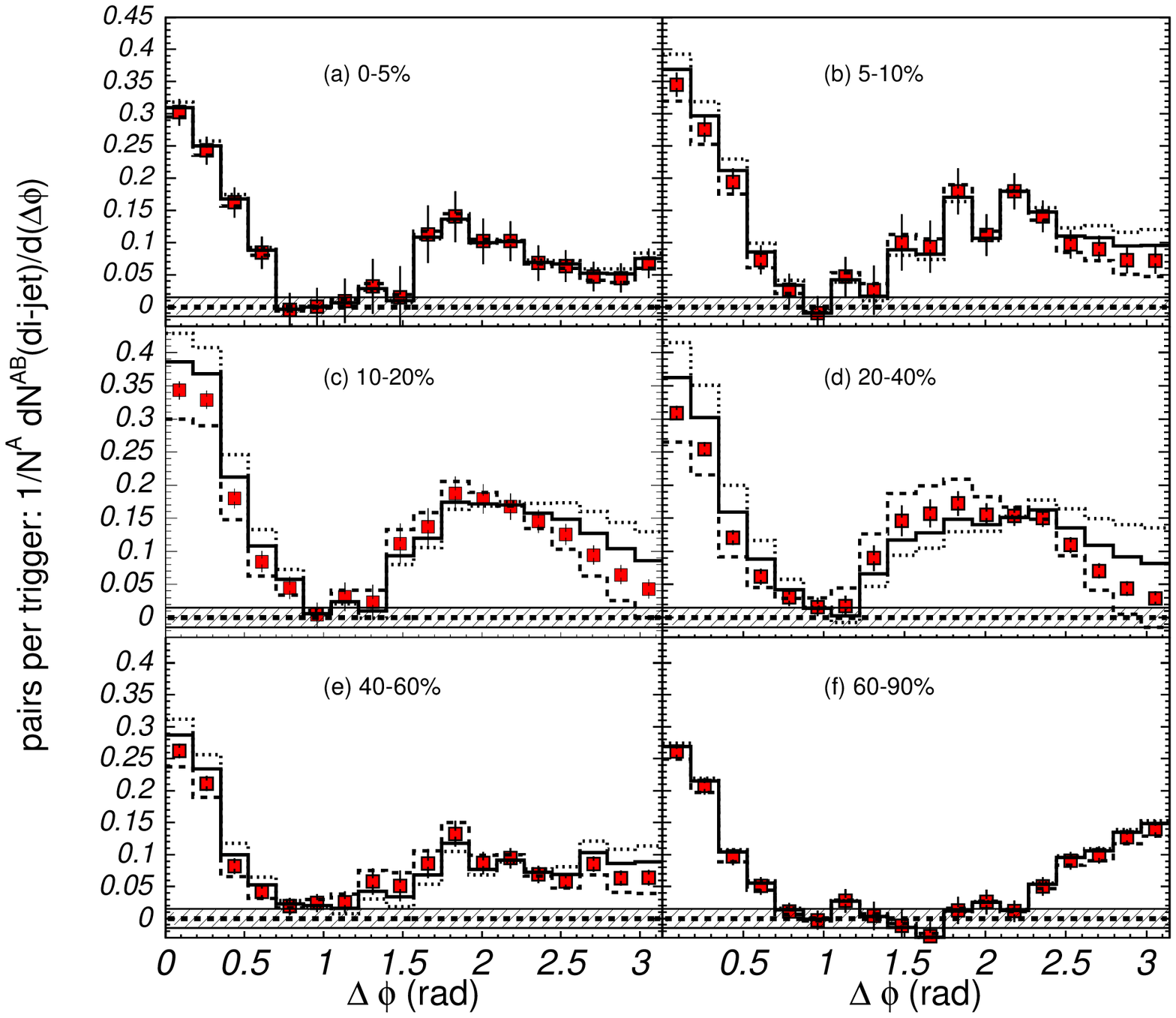,width=0.46\linewidth}
\end{tabular}
\end{center}\vspace*{-0.04\linewidth}
\caption[]{a)(left) Associated charged tracks with $2\leq p_{T_a}\leq p_{T_t}$ GeV/c per trigger charged particle with $6\leq p_{T_t}\leq 8$ GeV/c for central Au+Au collisions at $\sqrt{s_{NN}}=200$ GeV as a function of the angle $\Delta \phi$ between the tracks in the range $|\eta|<1.4$ compared to the data in p-p collisions added to the flow modulated Au+Au background~\cite{STARHardkeQM02}; b) (right) Associated charged tracks with $1\leq p_{T_a}\leq 2.5$ GeV/c per trigger charged particle with $2.5\leq p_{T_t}\leq 4.0$ GeV/c after subtraction of flow-modulated background. The dashed (solid) curves are the distributions that would result from increasing (decreasing) the flow modulation by one unit of the systematic error; the dotted curve would result from decreasing by two units~\cite{PXppg032}. Note that only the jet correlation, after background subtraction, is shown. \label{fig:jets}}
\end{figure}
In Fig.~\ref{fig:jets}a, the conditional probability of finding an associated charged particle with $2\leq p_{T_a}\leq p_{T_t}$ per trigger charged particle with $6\leq p_{T_t}\leq 8$ GeV/c is shown for p-p and Au+Au central collisions, where the p-p data have been added to the large flow-modulated Au+Au background. (Note the offset zero.) For the p-p data there two peaks, a same-side peak at $\Delta\phi=0$ where associated particles from the jet cluster around the trigger particle and a peak at $\Delta\phi=\pi$ radians, from the away jet. For Au+Au central collisions, the same side peak is virtually identical to that in p-p collisions, while the away peak, if any, is masked by the $v_2$ modulation, and, in any case, is much smaller than observed in p-p collisions~\cite{STARHardkeQM02}. The  `vanishing' of the away jet is consistent with jet quenching in the medium due to energy loss---the away parton loses energy, and perhaps stops, so that there are fewer fragments in a given $p_T$ range.  The fact that the number of associated particles in a cone around the trigger particle is the same in Au+Au as in p-p collisions is a strong argument against hadronic absorption as the cause of jet quenching~\cite{Cassing04}. Since all hadrons would be absorbed roughly equally, the associated peak would be suppressed as much as the inclusive spectrum in a hadronic scenario, which is clearly not seen. The only escape from this conclusion is if the partons or hadrons were so strongly absorbed in the medium that only jets emitted from the surface were seen. Of course, since pions at mid-rapidity with $p_T > 1.4$ GeV/c, $\gamma_T > 10$, can not be resolved before 14fm ($=\gamma_T\beta_T/m_0$), due to the uncertainty principle, they are formed by fragmentation  outside the medium,  even taking account of the flow velocity. Thus hadronic absorption in the medium is not possible for pions.  

    The away jet reappears if the transverse momentum of the associated charged particles is lowered to the range $1\leq p_{T_a}\leq 2.5$ GeV/c for trigger charged particles  with $p_{T_t}$ in the range $2.5\leq p_{T_t}\leq 4.0$ GeV/c (Fig.\ref{fig:jets}b)\cite{PXppg032,seealsoFQW}. In the most peripheral collisions,  the shape of the trigger and away jets looks the same as in p-p collisions (Fig.~\ref{fig:jets}a). However with increasing centrality, the away jet becomes much wider (Fig.~\ref{fig:jets}b) and possibly develops a dip at $\Delta \phi=\pi$. Since the outgoing partons travel much faster than the speed of sound in the medium, it has been proposed that a sonic-boom or mach-cone might develop, as suggested by the dip~\cite{ShuryakMIT}. There are many other ideas to explain the apparent dip, not the least of which is to get a better understanding of how exactly to extract the flow effect. 
    
   Study of jet correlations in A+A collisions is much more complicated than the same subject in p-p collisions and one can expect a long learning curve.   The next step in these studies is to measure jet suppression and correlations as a function of the angle $(\Delta\phi)$ to the reaction plane and centrality in an attempt to separate the effects of the density of the medium and the path length traversed. For a given centrality, variation of $\Delta\phi$ gives a variation of the path-length traversed for fixed initial conditions, while varying the centrality allows the initial conditions to vary. Although similar in spirit to a $v_2$ measurement, this is different in detail since measurement of $R_{AA}(\Delta\phi)$ is an absolute measurement, while $v_2$ is a relative measurement, and a first look has already yielded interesting results~\cite{PXppg054}.

\subsection{The smoking gun?}
 
   The jet suppression observed at RHIC is unique in that it had never been seen in either p+A collisions or in A+A collisions at lower $\sqrt{s_{NN}}$ and it probes the color charge density of the medium. Many questions and unsolved problems remain which are under active investigation, but this effect comes closest of all, in the author's opinion, to meeting the criteria for declaring the medium a Quark Gluon Plasma: 
   \begin{itemize}
   \item There is no such effect in p+A collisions at any $\sqrt{s_{NN}}$
\item It is not the `ordinary physics' of A+A collisons since it only occurs for $\sqrt{s_{NN}} \geq 30$ GeV
\item In all discussions of the effect, the operative `charge' is color and the operative degrees of freedom are quarks and gluons.
   \end{itemize}
  
\section{$J/\Psi$ Suppression}

\begin{figure}[!htb]
\begin{center}
\begin{tabular}{cc}
\psfig{file=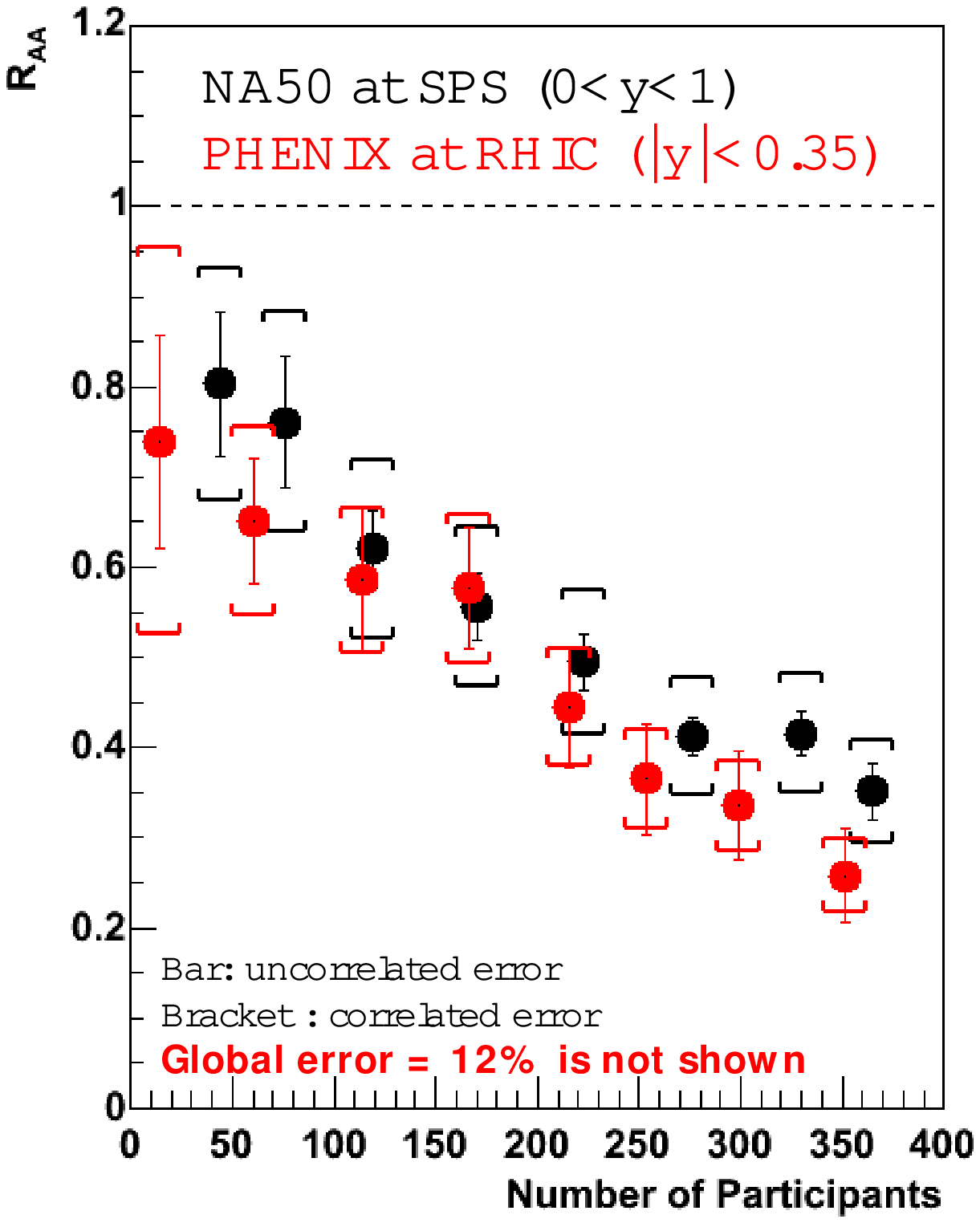,width=0.43\linewidth}&
\hspace*{-0.02\linewidth}\psfig{file=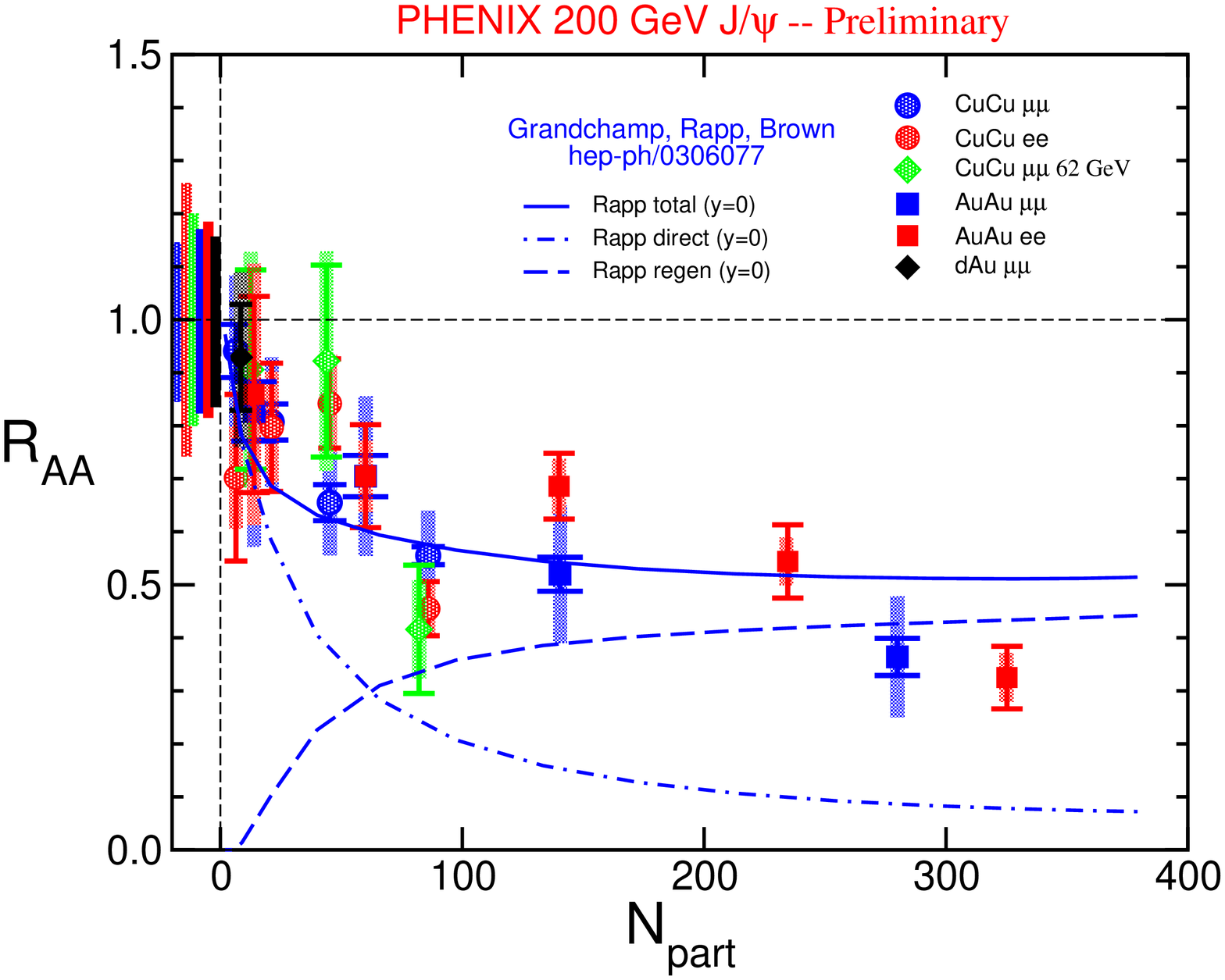,width=0.54\linewidth,height=0.59\linewidth}
\end{tabular}
\end{center}\vspace*{-0.25in}
\caption[]{$J/\Psi$ suppression relative to p-p collisions ($R_{AA}$) as a function of centrality ($N_{\rm part}$). a) (left) Measurements at RHIC~\cite{PXJPsiAuAu200,GunjiQM06} and at the CERN/SPS~\cite{NA50EPJC39}. b) (right) Comparison of RHIC data~\cite{GunjiQM06} from a) to predictions of a model~\cite{GrandchampPRL92} with (regen.) and without (direct) recombination.  \label{fig:NA50PX}}
\end{figure}

	Although $J/\Psi$ suppression was considered to be the `gold-plated' signature of the QGP and served as the inspiration for much theoretical and experimental work, the interpretation of the $J/\Psi$ suppression observed at the CERN/SPS~\cite{NA50PLB450,NA50EPJC39} in terms of a QGP was controversial, and the effect at the SPS could be explained by hadronic~\cite{CapellaFerreiro05,BKCS04} and even thermal models~\cite{GGPRL83}. The clincher would be the measurement of the $J/\Psi$ at RHIC. All the initially produced $J/\Psi$, the ones not suppressed at CERN, would be totally suppressed in the much hotter denser QGP at RHIC, which would prove that the $J/\Psi$ suppression at CERN was indeed the result of deconfinement. However, 
in the ensuing years a `nightmare scenario' developed when it was realized that if a QGP were indeed produced, the thermal $c,\bar{c}$ quarks would recombine~\cite{PBMStachelPLB490,ThewsPRC63} to form $J/\Psi$. Thus, if the $J/\Psi$ suppression were the same at RHIC as at CERN, this would imply that RHIC, not CERN, had discovered the QGP since all the initially produced $J/\Psi$, which were suppressed at CERN by whatever mechanism, would be totally suppressed at RHIC, leaving only the thermal $J/\Psi$ produced by recombination in a QGP.    The nightmare is that nobody would believe this explanation. Incredibly, this is exactly what happened (see Fig.~\ref{fig:NA50PX}), the $J/\Psi$ suppression, expressed as $R_{AA}$ turned out to be the same at RHIC~\cite{PXJPsiAuAu200} as in the famous SPS measurement~\cite{NA50EPJC39}. 

This effect is illustrated quantitatively in Fig.~\ref{fig:NA50PX}b. Models~\cite{SatzQM99,CapellaFerreiro05,BKCS04,GrandchampPRL92} which reproduce the SPS $J/\Psi$ suppression with or without a QGP predict a near total absence of $J/\Psi$ at RHIC beyond 150 participants without recombination. With recombination turned on, the RHIC data are reproduced, with one notable exception~\cite{AndronicPLB517} which predicts a larger $R_{AA}$ for $J/\Psi$ at RHIC than at the SPS. In fact, a $J/\Psi$ enhancement would have been the smoking gun for the QGP. It will be interesting to see whether this occurs at the LHC. 
    
    Does the agreement of the CERN/SPS and RHIC data for $R_{AA}$ in $J/\Psi$ production eliminate $J/\Psi$ suppression as a signature of deconfinement? Is it possible that the $J/\Psi$ ($c,\bar{c}$) is no different in its QGP sensitivity than the $\phi$ ($s,\bar{s}$)-meson? Recent increases in the predicted dissociation temperatures~\cite{Wong05}, give a possible way out~\cite{Satz05}. Satz has proposed  that the $\chi_c$ and the $\Psi^{'}$ were suppressed both at the SPS and at RHIC, but in neither place was the direct $J/\Psi$ suppressed, since a temperature $\sim 2T_c$ was not reached. However temperature sufficient to melt the $J/\Psi$ should be reached at the LHC. Does this mean that we have to wait until 2009 to prove or disprove an idea proposed in 1986? Fortunately there are other tests to be made on the RHIC data. If the $\chi_c$ or $\Psi^{'}$ were to be observed at RHIC, that would settle the issue. For recombination to be true, $J/\Psi$ flow should be observed~\cite{GrandchampPRL92}, and both the rapidity and the $p_T$ distributions of $J/\Psi$ should be much narrower due to recombination than for directly produced $J/\Psi$~\cite{ThewsMangano}. 
    
    One thing is perfectly clear from this discussion: the claim~\cite{HeinzJacob} of the QGP discovery from $J/\Psi$ suppression at the CERN/SPS was, at best, premature.   

\section{Conclusions}
   The medium produced at RHIC is not the expected gaseous Quark Gluon Plasma. It is a strongly interacting liquid of bare color charges, perhaps a perfect fluid. The medium flows and strongly absorbs `colored' objects, including embedded light and heavy quark probes from hard-scattering, quite unlike anything observed at lower c.m. energies. However the reaction of the medium to the $J/\Psi$ seems to be the same, in fact nearly identical, to the effect observed at lower $\sqrt{s_{NN}}$. There are also many other mysteries to be explained such as the baryon anomaly and the wide and apparently split jet-correlations, possibly indicative of a Mach-cone-like reaction of the medium. Nevertheless, it is clear that we have moved from period of discovery to a period of characterizing the properties of the medium which to quote my colleague Ed O'Brien, ``might exceed the proton in its importance to the understanding of non-perturbative QCD''.


\begin{thebibliography}{99}
\bibitem{specificity} It should be noted that the requirement of specificity to A+A collisions immediately rules out the QGP in p-p collisions, which is disputable. See Refs.~\refcite{Weiner05,Alexopoulos02}.
\end{thebibliography}
\end{document}